\documentclass[useAMS, usenatbib]{mn2e}
\usepackage{color}
\usepackage{epsfig}
\usepackage{epstopdf}
\usepackage{amsmath, amssymb}
\usepackage{hyperref}
\usepackage{wasysym} 

\newlength{\plotwidth}
\newlength{\fullwidth}
\renewcommand{\topmargin}{-1cm} 

\title[6\lowercase{d}FGS: Dependence of halo occupation on stellar mass]{The 6\lowercase{d}F Galaxy Survey: Dependence of halo occupation on stellar mass}
\author[Beutler et al.]
{\parbox{\textwidth}{Florian Beutler$^{1,2}$\thanks{E-mail: \texttt{fbeutler@lbl.gov}},
Chris Blake$^3$, Matthew Colless$^4$, D. Heath Jones$^5$,\\
Lister Staveley-Smith$^{2,6}$, Lachlan Campbell$^7$, Quentin Parker$^{4,8}$, Will Saunders$^4$, Fred Watson$^4$}\vspace{0.4cm}\\
\parbox{\textwidth}{
$^{1}$Lawrence Berkeley National Laboratory, 1 Cyclotron Road, Berkeley, CA 94720, USA\\
$^{2}$International Centre for Radio Astronomy Research (ICRAR), University of Western Australia, 35 Stirling Highway, Crawley WA 6009, Australia\\
$^{3}$Centre for Astrophysics \& Supercomputing, Swinburne University of Technology, P.O. Box 218, Hawthorn, VIC 3122, Australia\\
$^{4}$Australian Astronomical Observatory, PO Box 296, Epping NSW 1710, Australia\\
$^{5}$School of Physics, Monash University, Clayton, VIC 3800, Australia\\
$^{6}$ARC Centre of Excellence for All-sky Astrophysics (CAASTRO)\\
$^{7}$Western Kentucky University, Bowling Green, KY 42101, USA\\
$^{8}$Department of Physics and Astronomy, Faculty of Sciences, Macquarie University, NSW 2109, Sydney, Australia}}

\begin{document}

\label{firstpage}

\maketitle

\begin{abstract}
In this paper we study the stellar-mass dependence of galaxy clustering in the 6dF Galaxy Survey. The near-infrared selection of 6dFGS allows more reliable stellar mass estimates compared to optical bands used in other galaxy surveys. Using the Halo Occupation Distribution (HOD) model, we investigate the trend of dark matter halo mass and satellite fraction with stellar mass by measuring the projected correlation function, $w_p(r_p)$. We find that the typical halo mass ($M_1$) as well as the satellite power law index ($\alpha$) increase with stellar mass. This indicates, (1) that galaxies with higher stellar mass sit in more massive dark matter halos and (2) that these more massive dark matter halos accumulate satellites faster with growing mass compared to halos occupied by low stellar mass galaxies. 
Furthermore we find a relation between $M_1$ and the minimum dark matter halo mass ($M_{\rm min}$) of $M_1 \approx 22\,M_{\rm min}$, in agreement with similar findings for SDSS galaxies. The satellite fraction of 6dFGS galaxies declines with increasing stellar mass from $21\%$ at $M_{\rm stellar} = 2.6\times10^{10}h^{-2}\,M_{\odot}$ to $12\%$ at $M_{\rm stellar} = 5.4\times10^{10}h^{-2}\,M_{\odot}$ indicating that high stellar mass galaxies are more likely to be central galaxies. We compare our results to two different semi-analytic models derived from the Millennium Simulation, finding some disagreement. Our results can be used for placing new constraints on semi-analytic models in the future, particularly the behaviour of luminous red satellites. 
Finally we compare our results to studies of halo occupation using galaxy-galaxy weak lensing. We find good overall agreement, representing a valuable crosscheck for these two different tools of studying the matter distribution in the Universe.
\end{abstract}


\section{Introduction}
\label{sec:intro}
 
The first statistical studies of galaxy clustering~\citep{Totsuji:1969,Peebles:1973,Hauser:1973,Hauser:1974,Peebles:1974} found that the galaxy correlation function behaves like a power law, which is difficult to explain from first principles~\citep{Berlind:2001xk}. More recent studies, however, found deviations from a power law. For example~\citet{Zehavi:2004zn} showed that the projected correlation function $w_p(r_p)$ of SDSS galaxies exhibits a statistically significant departure from a power law. They also showed that a 3-parameter Halo Occupation Distribution (HOD) model (e.g.,~\citealt{Jing:1997nb, Ma:2000ik, Peacock:2000qk, Seljak:2000gq, Scoccimarro:2001, Berlind:2001xk, Cooray:2002dia}) together with a $\Lambda$CDM background cosmology, can account for this departure, reproducing the observed $w_p(r_p)$. 

Within the halo model the transition from the 1-halo term to the 2-halo term causes a ''dip'' in the correlation function at around $1-3h^{-1}\,$Mpc, corresponding to the exponential cutoff in the halo mass function. In case of a smooth transition between the one- and two-halo terms, this can mimic a power-law correlation function. Studies with Luminous Red Galaxies (LRGs) found that the deviation from a power-law is larger for highly clustered bright galaxies~\citep{Zehavi:2004zn, Zehavi:2004ii, Blake:2007xp, Zheng:2008np, Zehavi:2010bh}, and at high redshift~\citep{Conroy:2005aq}, which agrees with theoretical predictions~\citep{Watson:2011cz}.

While galaxy clustering is difficult to predict, dark matter clustering is dominated by gravity and can be predicted for a given cosmology using N-body simulations. Using models for how galaxies populate dark matter halos, which are usually motivated by N-body simulations, we can directly link galaxy clustering and matter clustering. This can be modelled in terms of the probability distribution $p(N|M)$ that a halo of virial mass $M$ contains $N$ galaxies of a given type. 
On strongly non-linear scales the dark matter distribution is given by the actual density distribution of the virialized halos, while on large and close to linear scales the dark matter distribution can be predicted from linear perturbation theory.

HOD modelling has been applied to galaxy clustering data from the 2-degree Field Galaxy Redshift Survey (2dFGRS)~\citep{Porciani:2004vi, Tinker:2006sk} and the Sloan Digital Sky Survey (SDSS)~\citep{vandenBosch:2002zn, Magliocchetti:2003ee, Zehavi:2004zn, Zehavi:2004ii, Tinker:2004gf, Yang:2004qi, Yang:2007pg, Zehavi:2010bh}. More recently it also became possible to model the clustering of high-z galaxies using VVDS~\citep{Abbas:2010hr}, Bo\"{o}tes~\citep{Brown:2008eb}, DEEP2~\citep{Coil:2005ku} and Lyman-break galaxies at high redshift in the GOODS survey~\citep{Lee:2005jha}. Such studies revealed, that the minimum mass, $M_{\rm min}$ for a halo to host a central galaxy  more luminous than some threshold, $L$ is proportional to $L$ at low luminosities, but steepens above $L_*$. Massive halos have red central galaxies with predominantly red satellites, while the fraction of blue central galaxies increases with decreasing host halo mass. Furthermore~\citet{Zehavi:2004ii} found that there is a scaling relation between the minimum mass of the host halos, $M_{\rm min}$ and the mass scale, $M_1$ of halos that on average host one satellite galaxy in addition to the central galaxy, $M_1 \approx 23\,M_{\rm min}$. Using a different HOD parameterization, \citet{Zheng:2007zg} found the relation to be $M_1 \approx 18\,M_{\rm min}$, very similar to~\citet{Zehavi:2010bh} who found $M_1 \approx 17\,M_{\rm min}$. 

The 6dF Galaxy Survey is one of the biggest galaxy surveys available today with a sky coverage of $42\%$ and an average redshift of $\overline{z} = 0.05$. The survey includes about $125\,000$ redshifts selected in the $J,H,K,b_J,r_F$-bands~\citep{Jones:2004zy,Jones:2005ya,Jones:2009yz}. The near infrared selection, the high completeness and the wide sky coverage make 6dFGS one of the best  surveys in the local Universe to study galaxy formation. This dataset has been used to study the large scale galaxy clustering to measure the Hubble constant using Baryon Acoustic Oscillations~\citep{Beutler:2011hx}, as well as the growth of structure at low redshift~\citep{Beutler:2012px}. While these previous studies used the $K$-band selected sample, in this analysis we use the $J$-band. The $J$-band allows the most reliable stellar mass estimate of the five bands available in 6dFGS, because of its lower background noise. Together with the $b_J-r_F$ colour we can derive stellar masses using the technique of~\citet{Bell:2000jt}, which leads to a dataset of $76\,833$ galaxies in total. The photometric near-infrared selection from 2MASS makes the stellar mass estimates in 6dFGS more reliable than stellar mass estimates in other large galaxy surveys which rely on optical bands~\citep{Drory:2004ib,Kannappan:2007ys,Longhetti:2008gv,Grillo:2007kg,Gallazzi:2009aj}.

Numerical N-body simulations are usually restricted to dark matter only. To understand galaxy formation, baryonic effects such as feedback and gas cooling, have to be included. Such simulations face severe theoretical and numerical challenges. Semi-analytic models build upon pre-calculated dark matter merger trees from cosmological simulations and include simplified, physically and observationally motivated, analytic recipes for different baryonic effects. Semi-analytic models have been shown to successfully reproduce observed statistical properties of galaxies over a large range of galaxy masses and redshifts (e.g.~\citealt{Croton:2005fe,Bower:2005vb,De Lucia:2006vua,Bertone:2007sj,Font:2008pc,Guo:2009fn}) and allow a level of understanding unavailable in N-body simulations. The underlying models are necessarily simplified and often use a large number of free parameters to fit different observations simultaneously. In this paper, we derive 6dFGS mock surveys from semi-analytic models based on the Millennium Simulation~\citep{Springel:2005nw} and compare the properties of these surveys with measurements in 6dFGS. Our results can be used to improve upon these semi-analytic models and further our understanding of baryonic feedback processes on galaxy clustering.

\citet{Mandelbaum:2005nx} studied halo occupation as a function of stellar mass and galaxy type using galaxy-galaxy weak lensing in SDSS. They found that for a given stellar mass, the halo mass is independent of morphology below $M_{\rm stellar} = 10^{11}\,M_{\odot}$, indicating that stellar mass is a good proxy for halo mass at that range. We compare our results with~\citet{Mandelbaum:2005nx} which represents a valuable crosscheck of the HOD analysis using two very different techniques, weak lensing and galaxy clustering. 

This paper is organised as follows: First we introduce the 6dF Galaxy Survey in section~\ref{sec:survey} together with the technique to derive the stellar masses. We also explain how we derive our four volume-limited sub-samples in stellar mass and redshift, which are then used for the further analysis. In section~\ref{sec:analysis} we calculate the projected correlation function, $w_p(r_p)$, for each sub-sample and use jack-knife re-sampling to derive the covariance matrices. In section~\ref{sec:pl} we fit power laws to the projected correlation functions of the four sub-samples. In section~\ref{sec:hod} we introduce the HOD framework and in section~\ref{sec:hodfits} we apply the HOD model to the data. In section~\ref{sec:sam} we derive 6dFGS mock samples from two different semi-analytic models, which we then compare to our results in section~\ref{sec:results} together with a general discussion of our findings. We conclude in section~\ref{sec:con}.

Throughout the paper we use $r$ to denote real space separations and $s$ to denote separations in redshift space. Our fiducial model to convert redshifts into distances is a flat universe with $\Omega_m = 0.27$, $w = -1$ and $\Omega_k = 0$. The Hubble constant is set to $H_0 = 100h\,$km\,s$^{-1}$Mpc$^{-1}$ which sets the unit of stellar masses to $h^{-2}M_{\odot}$, while most other masses are given in $h^{-1}M_{\odot}$. The HOD model uses cosmological parameters following WMAP7~\citep{Komatsu:2010fb}.

\section{The 6dF Galaxy survey}
\label{sec:survey}

The 6dF Galaxy Survey~\citep[6dFGS;][]{Jones:2004zy,Jones:2005ya,Jones:2009yz} is a near-infrared selected ($J,H,K$) redshift survey covering $17\,000\,$deg$^2$ of the southern sky. The $J$, $H$ and $K$ surveys avoids a $\pm10^\circ$ region around the Galactic Plane to minimise Galactic extinction and foreground source confusion in the Plane. The near-infrared photometric selection was based on total magnitudes from the Two-Micron All-Sky Survey Extended Source Catalog~\citep[2MASS XSC;][]{Jarrett:2000me}. The spectroscopic redshifts of 6dFGS were obtained with the Six-Degree Field (6dF) multi-object spectrograph of the UK Schmidt Telescope (UKST) between 2001 and 2006.

The 6dFGS $J$-selected sample used in this paper contains (after completeness cuts) $76\,833$ galaxies selected with $9.8 \leq J \leq 13.75$. We chose the $J$-band because it has the highest signal-to-noise of the three 2MASS bands. While there is slightly less extinction in the $K$-band compared to the $J$-band, for practical purposes, the $J$-band has better S/N because the night sky background glow is much less in the $J$- than in the $K$-band. The near infrared selection makes 6dFGS very reliable for stellar mass estimates.

The mean completeness of 6dFGS is $92$ percent and the median redshift is $z = 0.05$. Completeness corrections are derived by normalising completeness-apparent magnitude functions so that, when integrated over all magnitudes, they equal the measured total completeness on a particular patch of sky. This procedure is outlined in the luminosity function evaluation of~\citet{Jones:2006xy} and also in Jones et al., (in prep). The original survey papers~\citep{Jones:2004zy,Jones:2005ya,Jones:2009yz} describe in full detail the implementation of the survey and its associated online database.

The clustering in a galaxy survey is estimated relative to a random (unclustered) distribution which follows the same angular and redshift selection function as the galaxy sample itself. We base our random mock catalogue generation on the 6dFGS luminosity function~\citep{Jones:2006xy}, where we use random numbers to pick volume-weighted redshifts and luminosity function-weighted absolute magnitudes. We then test whether the redshift-magnitude combination falls within the 6dFGS $J$-band faint and bright apparent magnitude limits ($9.8 \leq J \leq 13.75$). We assigned a $b_J$-$r_F$ colour to each random galaxy using the redshift- $b_J$-$r_F$ colour relation measured in the data and used these to derive stellar masses for the random galaxies using the same technique as for the actual galaxies (see section~\ref{sec:stellar}).

\subsection{Stellar mass estimate and volume-limited sub-samples}
\label{sec:stellar}

To calculate the stellar mass for our dataset we use the stellar population synthesis results from~\citet{Bruzual A.:1993is} together with a scaled Salpeter initial mass function (IMF) as reported in~\citet{Bell:2000jt}
\begin{equation}
\begin{split}
\log_{10}(M_{\rm stellar}/L_J) &= -0.57C_{b_J-r_F} + 0.48\\
\log_{10}(L_J) &= (M_J^{\rm sun} - M_J)/2.5\\
\log_{10}(M_{\rm stellar}) &= \log_{10}(M_{\rm stellar}/L_J) + \log_{10}(L_J),
\label{eq:stellar}
\end{split}
\end{equation}
with the 2MASS $b_J$-$r_F$ colours, $C_{b_J-r_F}$ the $J$-band absolute magnitude, $M_J$ and the $J$-band absolute magnitude of the sun, $M_J^{\rm sun} = 3.70$~\citep{Worthey:1994iw}. The biggest uncertainty in stellar mass estimates of this type is the choice of the IMF. Assuming no trend in IMF with galaxy type, the range of IMFs presented in the literature cause uncertainties in the absolute normalisation of the stellar $M/L$-ratio of a factor of $2$ in the near-infrared~\citep{Bell:2000jt}. The 6dFGS stellar mass function as well as a comparison of different stellar mass estimates is currently in preparation (Jones et al. in prep).

\begin{figure}
\begin{center}
\epsfig{file=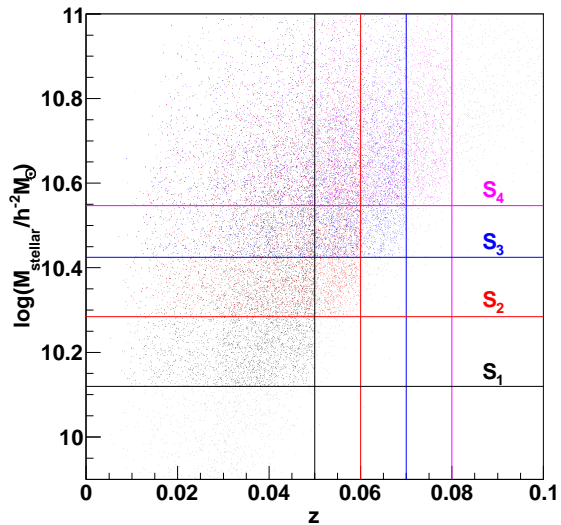,width=8cm}
\caption{The distribution of 6dFGS galaxies in log stellar mass and redshift. The redshift and stellar mass cuts imposed to create the four volume-limited samples ($S_1$ to $S_4$ see Table~\ref{tab:vls}) are shown by the coloured lines. All galaxies in the upper left quadrant created by the two correspondingly coloured lines are included in the volume-limited sub-samples. The plot shows a randomly chosen set of $20\%$ of all galaxies.}
\label{fig:z_vs_logSM2}
\end{center}
\end{figure}

\begin{figure}
\begin{center}
\epsfig{file=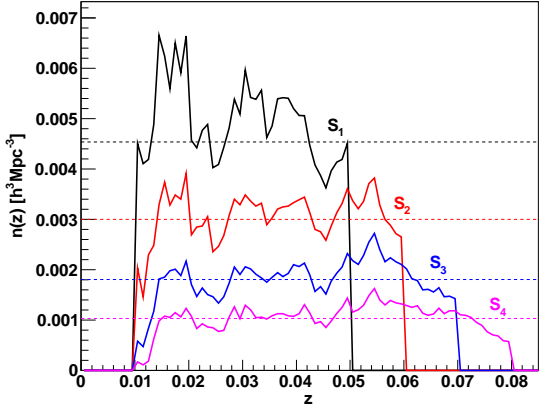,width=8cm}
\caption{Galaxy density as a function of redshift for the four different volume-limited sub-samples $S_1$-$S_4$. All samples follow an approximately constant number density indicated by the dashed lines and listed in the last column of Table~\ref{tab:vls}.}
\label{fig:nz}
\end{center}
\end{figure}

We create four volume-limited sub-samples in redshift and stellar mass. This is done by choosing an upper limit in redshift ($z_{\rm max}$) and then maximising the number of galaxies by choosing a lower limit in stellar mass ($M^{\rm min}_{\rm stellar}$), meaning that every galaxy above that stellar mass will be detected in 6dFGS, if its redshift is below the redshift limit (see Figure~\ref{fig:z_vs_logSM2}). Because of the distribution in $b_J$-$r_F$ colour, a clear cut in absolute magnitude does not correspond to a clear cut in stellar mass and therefore our samples are not perfectly volume-limited. We use the absolute magnitude limit which corresponds to a chosen redshift limit and derive a stellar mass limit, using the $b_J$-$r_F$ colour corresponding to the $50\%$ height of the $b_J$-$r_F$ distribution ($b_J$-$r_F$($50\%$) $= 1.28$). We create four sub-samples ($S_1$-$S_4$) with upper redshift cuts at $z_{\rm max} = 0.05$, $0.06$, $0.07$ and $0.08$ and with the mean log stellar masses ranging from $\log_{10}(M_{\rm stellar}/h^{-2}\,M_{\odot}) = 10.41$ to $10.73$. We also include a low redshift cut-off at $z_{\rm min} = 0.01$. Since the stellar mass distributions for these samples overlap, especially for the higher stellar mass samples, the results are correlated to some extent (see Figure~\ref{fig:z_vs_logSM2}). All our sub-samples are summarised in Table~\ref{tab:vls}. Figure~\ref{fig:nz} shows the galaxy density as a function of redshift for the four sub-samples. The roughly constant number density with redshift shows that our sub-samples are close to volume-limited.

\begin{table*}
\begin{center}
\caption{Summary of the different volume-limited sub-samples, $S_1$-$S_4$, used in this analysis. The effective redshift, $z_{\rm eff}$ and effective stellar mass are calculated as the mean of all galaxy pairs contributing to correlation function bins between $0.1$ and $40h^{-1}\,$Mpc and the error is the standard deviation from the mean. The galaxy density $n_g$ is calculated as the number of galaxies $N$ divided by the co-moving sample volume, which is calculated from the maximum redshift $z_{\rm max}$.}
\begin{tabular}{ccccccc}
\hline
sample & $z_{\rm max}$ & $\log_{10}\left(\frac{M^{\rm min}_{\rm stellar}}{h^{-2}M_{\odot}}\right)$ & $\langle z\rangle$ & $\log_{10}\left(\frac{\langle M_{\rm stellar}\rangle}{h^{-2}M_{\odot}}\right)$ & N & $n_g$ $[h^{3}\text{Mpc}^{-3}]$\\
\hline
$S_1$ & $0.05$ & $10.12$ & $0.0369\pm0.0010$ & $10.410\pm0.010$ & $24\,644$ & $4.536\times 10^{-3}$\\
$S_2$ & $0.06$ & $10.28$ & $0.0453\pm0.0011$ & $10.532\pm0.010$ & $27\,999$ & $3.001\times 10^{-3}$\\
$S_3$ & $0.07$ & $10.42$ & $0.0521\pm0.0012$ & $10.635\pm0.018$ & $26\,584$ & $1.806\times 10^{-3}$\\
$S_4$ & $0.08$ & $10.55$ & $0.0585\pm0.0020$ & $10.730\pm0.013$ & $22\,497$ & $1.030\times 10^{-3}$\\
\hline
\hline
\end{tabular}
\label{tab:vls}
\end{center}
\end{table*}

\section{Data analysis}
\label{sec:analysis}


We measure the separation between galaxies in our survey along the line of sight ($\pi$) and perpendicular to the line of sight ($r_p$) and count the number of galaxy pairs on this two-dimensional grid. We do this for the 6dFGS data catalogue, a random catalogue with the same selection function, and a combination of data-random pairs. We call the pair-separation distributions obtained from this analysis step $DD(r_p,\pi)$, $RR(r_p,\pi)$ and $DR(r_p,\pi)$, respectively. In the analysis we used $30$ random catalogues with the same size as the real data catalogue and average $DR(r_p,\pi)$ and $RR(r_p,\pi)$. The random mocks are sampled from the 6dFGS luminosity function (\citet{Jones:2006xy} and also Jones et al, in prep.), and hence they contain the same evolution of luminosity with redshift that we see in 6dFGS itself. The redshift-space correlation function is then given by the~\citet{Landy:1993yu} estimator:
\begin{equation}
\xi(r_p,\pi) = 1 + \frac{DD(r_p,\pi)}{RR(r_p,\pi)} \left(\frac{n_r}{n_d} \right)^2 - 2\frac{DR(r_p,\pi)}{RR(r_p,\pi)} \left(\frac{n_r}{n_d} \right),
\label{eq:LS2}
\end{equation}
where the ratio $n_r/n_d$ is given by
\begin{equation}
\frac{n_r}{n_d} = \frac{\sum^{N_r}_iw_i}{\sum^{N_d}_jw_j}
\end{equation}
and the sums go over all random ($N_r$) and data ($N_d$) galaxies. Here we employ a completeness weighting, $w_i$, where we weight each galaxy by the inverse sky- and magnitude completeness at its area of the sky~(Jones et al., in prep.). 

From the two-dimensional correlation function, $\xi(r_p,\pi)$, we calculate the projected correlation function
\begin{equation}
w_p(r_p) = 2\int^{\pi^{\rm max}}_0d\pi\, \xi(r_p,\pi),
\label{eq:wp}
\end{equation}
where we bin $\xi(r_p,\pi)$ in $30$ logarithmic bins from $0.1$ to $100h^{-1}\,$Mpc in $r_p$ and $\pi$. The upper integration limit in eq.~\ref{eq:wp} was chosen to be $\pi^{\rm max} = 50h^{-1}$Mpc for all sub-samples. We tested different values for $\pi^{\rm max}$, changing it between $20$ and $90h^{-1}\,$Mpc without  any significant effect to $r_0$ or $\gamma$ if the errors on $w_p(r_p)$ are adjusted accordingly. The error on scales larger than $r_p = 20$ is very large and hence the contribution of these scales to the fit is small.

To derive a covariance matrix for the projected correlation function we use the method of jack-knife re-sampling our galaxy samples. First we divide the dataset into $N = 400$ subsets, selected in R.A. and Dec. Each re-sampling step excludes one subset before calculating the correlation function. The covariance matrix is then given by
\begin{equation}
C_{ij} = \frac{(N-1)}{N}\sum^N_{k=1}\left[w^k(r_p^i) - \overline{w}(r_p^i)\right]\left[w^k(r_p^j) - \overline{w}(r_p^j)\right],
\end{equation}
where $w^k(r_p^i)$ is the projected correlation function estimate at separation $r_p^i$ with the exclusion of subset $k$ and $\overline{w}(r_p^i)$ is the mean. 

We also note that wide-angle effects can be neglected in this analysis, since we are interested in small-scale clustering (see~\citet{Beutler:2011hx} and~\citet{Beutler:2012px} for a detailed investigation of wide-angle effects in 6dFGS). 

\subsection{Fibre proximity limitations}
\label{sec:fibre}

The design of the 6dF instrument does not allow fibres to be placed closer than $5.7\,$arcmin~\citep{Jones:2004zy}, which corresponds to a distance of $r_p \approx 0.3h^{-1}\,$Mpc at redshift $z = 0.07$. This limitation is relaxed in 6dFGS where about $70\%$ of the survey area has been observed multiple times. However, for the remaining $30\%$ we have to expect to miss galaxy pairs with a separation smaller than $5.7\,$arcmin.

\begin{figure}
\begin{center}
\epsfig{file=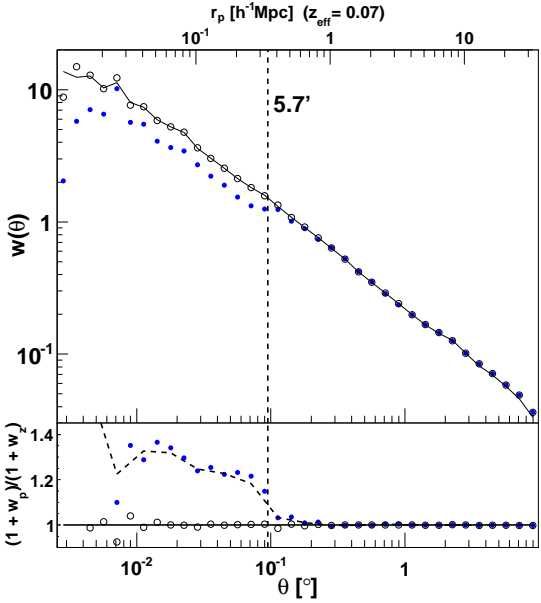,width=8cm}
\caption{The angular correlation function of 6dFGS (blue data points) as well as the target catalogue (black solid line). The deviation between the two angular correlation functions is caused by fibre proximity limitations. The lower panel shows the ratio of the two correlation functions, which can be used as a weight to correct for the fibre proximity limitations (see section~\ref{sec:fibre}). The open data points show the corrected 6dFGS angular correlation function which is in very good agreement with the target catalogue.}
\label{fig:fibre}
\end{center}
\end{figure}

Figure~\ref{fig:fibre} shows the angular correlation function for the 6dFGS redshift catalogue ($w_z$, blue data points) and the target catalogue ($w_p$, solid black line). While the target catalogue contains $104\,785$ galaxies, the redshift catalogue of galaxies which have a $J$-band, $b_J$-band and $r_F$-band magnitude contains $76\,833$. The dashed line indicates the angular scale of the fibres. The two angular correlation functions agree on scales $\theta > 0.1^\circ$, with the redshift catalogue falling below the target catalogue at lower scales.

In the lower panel of Figure~\ref{fig:fibre} we show the ratio $(1+w_p)/(1+w_z)$. The dashed line shows a spline fit to the blue data points, which than can be used to up-weight galaxy pairs with small angular separations and correct for the fibre proximity effect~\citep{Hawkins:2002sg}. This weighting is additional to the completeness weighting $w_i$ we introduced earlier. While the completeness weighting is applied to single data- as well as random galaxies, the fibre proximity weighting is only applied to data galaxy pairs. Applying this weight to the 6dFGS redshift catalogue results in the open data points in Figure~\ref{fig:fibre} which now agree very well with the target catalogue.

\section{Power law fits}
\label{sec:pl}

The projected correlation function can be related to the real-space correlation function $\xi(r)$, using~\citep{Davis:1982gc}
\begin{equation}
\begin{split}
w_p(r_p) &= 2\int^{\infty}_0dy\;\xi\left[(r_p^2 + y^2)^{1/2}\right]\\
&= 2\int^{\infty}_{r_p}r\,dr\,\xi(r)(r^2-r_p^2)^{-1/2}.
\end{split}
\label{eq:pro3}
\end{equation}
If the correlation function is assumed to follow a power law, $\xi(r) = (r/r_0)^{\gamma}$, with the clustering amplitude $r_0$ and the power law index $\gamma$, this can be written as
\begin{equation}
w_p(r_p) = r_p\left(\frac{r_p}{r_0}\right)^{-\gamma}\Gamma\left(\frac{1}{2}\right)\Gamma\left(\frac{\gamma-1}{2}\right)/\Gamma\left(\frac{\gamma}{2}\right),
\label{eq:pl}
\end{equation}
with $\Gamma$ being the Gamma-function. Using this equation we can infer the best-fit power law for $\xi(r)$ from $w_p(r_p)$. 

\begin{table}
\begin{center}
\caption{Best fitting parameters, $r_0$ and $\gamma$ for power law fits to the projected correlation functions $w_p(r_p)$ of the four volume-limited 6dFGS sub-samples ($S_1$-$S_4$). The fitting range in all cases is $0.1 < r_p < 40\,h^{-1}$Mpc with $24$ bins and $2$ free parameters. The last column shows the reduced $\chi^2$ indicating the goodness of the fit.}
\begin{tabular}{cccc}
\hline
sample & $r_0$ [$h^{-1}$Mpc] & $\gamma$ & $\chi^2/\rm d.o.f.$\\
\hline
$S_1$ & $5.14\pm0.23$ & $1.849\pm0.025$ & $23.5/(24-2) = 1.07$\\
$S_2$ & $5.76\pm0.17$ & $1.826\pm0.019$ & $19.5/(24-2) = 0.89$\\
$S_3$ & $5.76\pm0.16$ & $1.847\pm0.019$ & $32.6/(24-2) = 1.48$\\
$S_4$ & $6.21\pm0.17$ & $1.846\pm0.019$ & $35.8/(24-2) = 1.63$\\
\hline
\hline
\end{tabular}
\label{tab:pls}
\end{center}
\end{table}

\begin{figure*}
\begin{center}
\epsfig{file=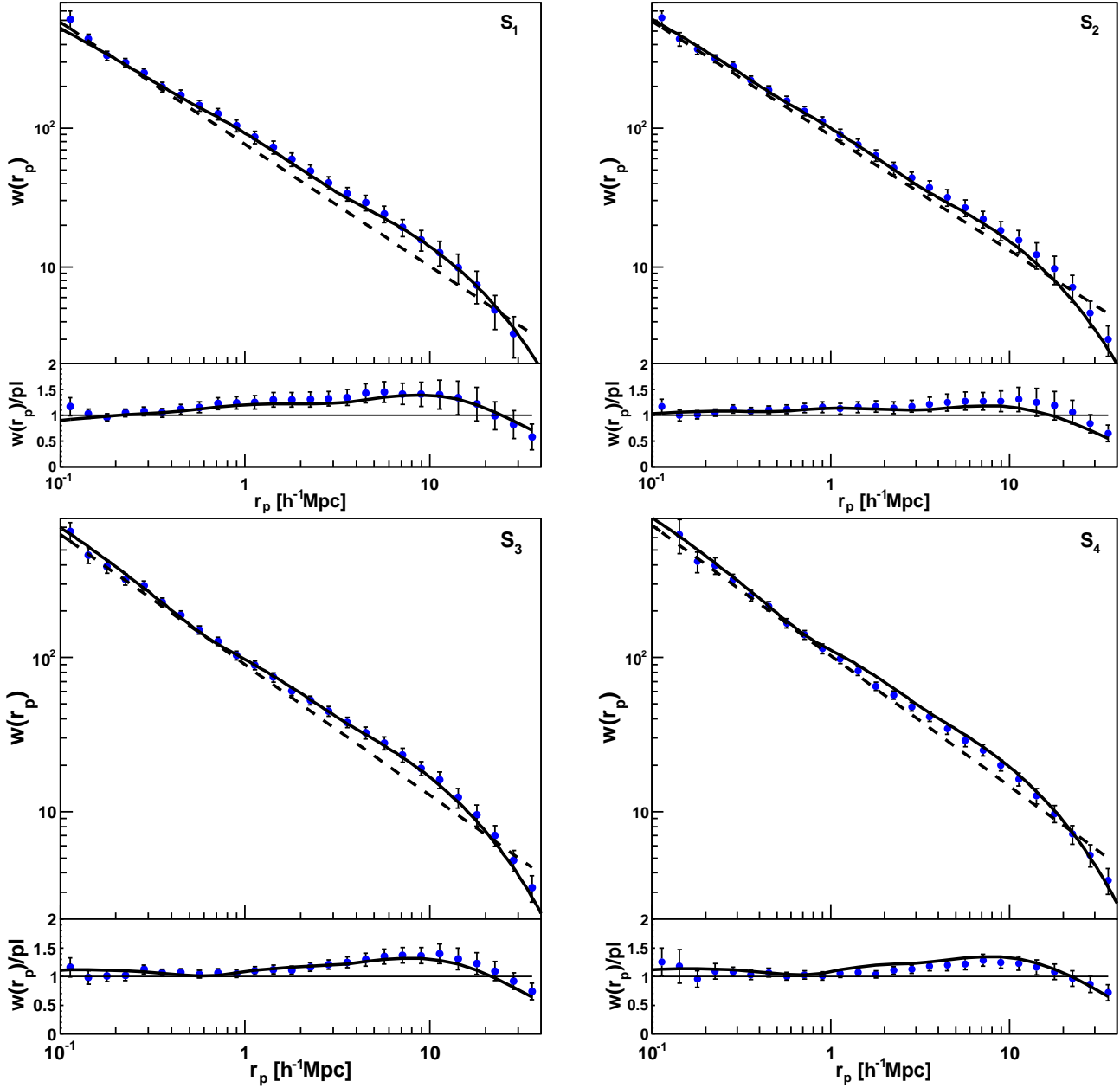,width=18cm}
\caption{The projected correlation function for the four different volume-limited 6dFGS sub-samples ($S_1$-$S_4$). The dashed black lines show the best fitting power laws (see Table~\ref{tab:pls}) while the solid black lines show the best fitting HOD models, derived by fitting the data between $0.1 \leq r_p \leq 40h^{-1}\,$Mpc. The lower panel shows the data and HOD models divided by the best fitting power laws.}
\label{fig:xi_hod_all}
\end{center}
\end{figure*}

Table~\ref{tab:pls} summarises the results of the power law fits to the four 6dFGS sub-samples. The best fitting power laws are also included in Figure~\ref{fig:xi_hod_all} together with the four projected correlation functions. We set the fitting range to be $0.1 < r_p < 40\,h^{-1}$Mpc which includes $24$ bins. Although 6dFGS has very good statistics at scales smaller than $0.1h^{-1}\,$Mpc we do not use them for our fits, since at such scales the fibre proximity correction becomes more than $30\%$ (see section~\ref{sec:fibre} and Figure~\ref{fig:fibre}). 

The value of the clustering amplitude, $r_0$, increases from $5.14$ to $6.21h^{-1}\,$Mpc with increasing stellar mass, while there doesn't seem to be a clear trend in $\gamma$, which varies around the value of $\gamma = 1.84$. The reduced $\chi^2$ in the last column of Table~\ref{tab:pls} indicates a good fit to the data for the first two sub-samples but grows to $\chi^2/\rm d.o.f. = 1.48$ and $1.63$ for the high stellar mass sub-samples, indicating deviations from a power law. The lower panels of Figure~\ref{fig:xi_hod_all} show the different projected correlation functions divided by the best fitting power law (blue data points). Here we can see that the deviations show systematic patterns. Such effects could be related to the strong correlations between bins in the correlation function. Nevertheless, these patterns can be addressed with a full HOD analysis, which we will pursue in the next section. We will compare the result of our power law fits with other studies in a more detailed discussion in section~\ref{sec:results}.

We remark at this point, that there is no satisfying theoretical model that predicts a power law behaviour of the correlation function, and hence the motivation of such a fit can only be empirical. While alternative approaches (e.g. the HOD model) rely on assumptions about the clustering behaviour of galaxies and dark matter, they are physically motivated and hence allow us to learn more about galaxy clustering than a pure empirical power law fit.

\section{Theory: Halo Occupation Distribution}
\label{sec:hod}

The Halo Occupation Distribution (HOD) model describes the relation between galaxies and mass in terms of the probability distribution $p(N|M)$ that a halo of virial mass $M$ contains $N$ galaxies of a given type. Knowing how galaxies populate dark matter halos, we can use a dark matter correlation function and infer the galaxy correlation function. We use CAMB~\citep{Lewis:2002ah} to derive a model matter power spectrum which we turn into a correlation function using a Hankel transform
\begin{equation}
\xi(r) = \frac{1}{2\pi^2}\int^{\infty}_0dkP(k)k^2\frac{\sin(kr)}{kr}.
\end{equation}
The underlying cosmological model is fixed to ($\Omega_bh^2$, $\Omega_ch^2$, $n_s$, $\sigma_8$) $=$ ($0.02227$, $0.1116$, $0.966$, $0.8$) as reported in~\citet{Komatsu:2010fb} (we set $\sigma_8=0.9$ for one special case). We have to be aware of the fact that the fitted HOD parameters depend somewhat on the assumed values of $\Omega_m$ and $\sigma_8$ and hence the absolute values of the HOD parameters could be biased, if the assumed cosmology is wrong. However in this study we focus on the relative HOD parameters for different stellar mass selected sub-samples, which is fairly robust against such uncertainties.

Here we employ an analytic HOD methodology that is similar to that of~\citet{Zheng:2007zg,Blake:2007xp} and~\citet{Zehavi:2010bh}. We utilise analytic approximations for the halo mass function~\citep{Tinker:2008ff}, the biased clustering of halos~\citep{Tinker:2004gf} and the nonlinear dark matter power spectrum~\citep{Smith:2002dz}. The profile of dark matter within halos is well described by the NFW profile~\citep{Navarro:1996gj} parameterized by the the concentration-mass relation~\citep{Duffy:2008pz}
\begin{equation}
c(M,z) = 6.71\left(\frac{M}{M_{\text{pivot}}}\right)^{-0.091}(1+z)^{-0.44}
\label{eq:duffy}
\end{equation}
with $M_{\text{pivot}} = 2\times10^{12}h^{-1}M_{\odot}$. We tried replacing eq.~\ref{eq:duffy} with the form suggested by~\citet{Bullock:1999he}, and found that the best-fitting HOD parameters changed by much less than the statistical errors.

\subsection{HOD framework and formalism}

In the HOD parametrisation it is common to separate the clustering contributions from the most massive galaxies, which are assumed to sit in the halo centre, from satellite galaxies. This picture of how galaxies populate halos is supported by hydrodynamic simulations (e.g.~\citealt{Berlind:2002rn, Simha:2008hd}) and semi-analytic models (e.g.~\citealt{White:1991mr,Kauffmann:1993gv,Croton:2005fe,Bower:2005vb}). The mean central and satellite number density of galaxies that populate dark matter halos of mass $M$ is~\citep{Zheng:2004id}
\begin{equation}
\begin{split}
\langle N_{c}(M)\rangle &= \begin{cases} 0 & \text{ if }\;M < M_{\rm min}\cr
								1 & \text{ if }\;M \geq M_{\rm min}\end{cases},\\
\langle N_{s}(M)\rangle &= \left(\frac{M}{M_1}\right)^{\alpha},
\label{eq:hod}
\end{split}
\end{equation}
where $M_{\rm min}$ is the minimum dark matter halo mass which can host a central galaxy, $M_1$ corresponds to the mass of halos that contain, on average, one additional satellite galaxy ($\langle N_s(M_1)\rangle = 1$) and $\alpha$ sets the rate at which halos accumulate satellites when growing in mass. In very massive halos the number of satellites is proportional to halo mass $M$ to the power of $\alpha$. The total HOD number is given by
\begin{equation}
\langle N_{t}(M)\rangle = \langle N_{c}(M)\rangle \left[1 + \langle N_{s}(M)\rangle \right],
\end{equation}
so that a dark matter halo can only host a satellite galaxy if it contains already a central galaxy. In our model we assume a step like transition from $\langle N_{c}(M)\rangle = 0$ to $\langle N_{c}(M)\rangle = 1$. In reality this is more likely to be a gradual transition with a certain width $\sigma_{\rm log M}$~\citep{More:2008za}. To account for this, other studies (e.g.~\citealt{Zehavi:2010bh}) modify the HOD parametrisation of the central halo term to
\begin{equation}
\langle N_{c}(M)\rangle = \frac{1}{2}\left[1 + \text{erf}\left(\frac{\log_{10}(M) - \log_{10}(M_{\rm min})}{\sigma_{\rm log M}}\right)\right],
\label{eq:Nc2}
\end{equation}
which turns into eq.~\ref{eq:hod} for the case $\sigma_{\rm log M} = 0$. For our data we found that the reduction in $\chi^2$ obtained from fits including $\sigma_{\rm log M}$ as an additional parameter is not big enough to justify this parameterisation. For our largest sub-sample (S2) we found $\Delta\chi^2 = -0.78$, while a new parameter would be justified when $\Delta\chi^2 < -2$ following the Akaike information criterion~\citep{Akaike:1974}. 

There are also higher parameter models for the satellite fraction such as
\begin{equation}
\langle N_{s}(M)\rangle = \left(\frac{M-M_0}{M_1}\right)^{\alpha},
\end{equation}
where $M_0$ is the minimum halo mass at which satellites can exist. Again we tested this model and found that the reduction in $\chi^2$ does not justify this additional parameter. Hence we chose the three parameter model ($M_1$, $\alpha$, $M_{\rm min}$) of eq~\ref{eq:hod}\footnote{$M_{\rm min}$ is fixed by the number density}. 

Within the halo model we can account separately for the clustering amplitude of galaxies which sit in the same dark matter halo (one halo term) and galaxies which sit in different dark matter halos (two halo term). At small scales the clustering will be dominated by the one halo term and at large scales it will be dominated by the two halo term. For the correlation function this can be written as
\begin{equation}
\xi(r) = \left[1 + \xi_{1h}(r)\right] + \xi_{2h}(r).
\end{equation}
where $\xi_{1h}(r)$ and $\xi_{2h}(r)$ represent the one halo and two halo terms respectively.

\subsection{The 1-halo term, $\xi_{1h}(r)$}

We separate the 1-halo term into contributions from central-satellite galaxy pairs and satellite-satellite galaxy pairs. The central-satellite contribution is given by
\begin{equation}
\xi_{1h}^{c-s}(r) = \frac{2}{n_g^2}\int^{\infty}_{M_{\rm vir}(r)} dM \frac{dn(M)}{dM}N_{c}(M)N_{s}(M)\frac{\rho(r,M)}{M},
\end{equation}
where $n_g$ is the galaxy number density and $\rho(r,M)$ is the halo density profile. The lower limit for the integral is the virial mass $M_{\rm vir}(r)$ corresponding to the virial separation $r_{\rm vir}$.\\
The satellite-satellite contribution is usually given by a convolution of the halo density profile with the halo mass function. Here we calculate this term in k-space since a convolution then turns into a simple multiplication
\begin{equation}
P_{1h}^{s-s}(k) = \frac{1}{n^2_g}\int^{\infty}_0dM \frac{dn(M)}{dM}N_{c}(M)N_{s}^2(M)\lambda(k,M)^2,
\label{eq:ssterm}
\end{equation} 
where $\lambda(k,M)$ is the normalised Fourier transform of the halo density profile $\rho(r,M)$. The real-space expression of eq.~(\ref{eq:ssterm}) is
\begin{equation}
\xi_{1h}^{s-s}(r) = \frac{1}{2\pi^2}\int^{\infty}_0dk\,P_{1h}^{s-s}(k)k^2\frac{\sin(kr)}{kr},
\end{equation}
which can then be combined with the central-satellite contribution to obtain  the 1-halo term
\begin{equation}
\xi_{1h}(r) = \xi_{1h}^{c-s}(r) + \xi_{1h}^{s-s}(r) - 1.
\end{equation}

\subsection{The 2-halo term, $\xi_{2h}(r)$}
\label{sec:twoh}

The 2-halo term, $\xi_{2h}(r)$, can be calculated from the dark matter correlation function since on sufficiently large scales the galaxy and matter correlation function are related by a constant bias parameter. We calculate the two halo term in Fourier space as
\begin{equation}
\begin{split}
P_{2h}(k,r) = &P_m(k)\times\\
              &\left[\int^{M_{\rm lim}(r)}_0 dM\frac{dn(M)}{dM}b_h(M,r)\frac{N_{t}(M)}{n'_g(r)}\lambda(k,M)\right]^2,
\end{split}
\label{eq:term2}
\end{equation}
where $P_m(k)$ is the non-linear model power spectrum from CAMB including halofit~\citep{Smith:2002dz} and $M_{\rm lim}(r)$ is the halo mass limit for which we can find galaxy pairs with a separation larger than $r_{\rm vir}$. To fix the value of $M_{\rm lim}(r)$ we use the number density matching model in the appendix of~\citet{Tinker:2005na}. $b_h(M,r)$ is the scale dependent halo bias at separation $r$ for which we assume the following model~\citep{Tinker:2004gf}
\begin{equation}
b_h^2(M,r) = b^2(M)\frac{\left[1+1.17\xi_m(r)\right]^{1.49}}{\left[1+0.69\xi_m(r)\right]^{2.09}},
\end{equation}
where $\xi_m(r)$ is the non-linear matter correlation function and $b(M)$ being the bias function of~\citet{Tinker:2004gf}.

\subsection{Derived quantities}
\label{sec:derived}

Since the HOD model is directly based on a description of dark matter clustering and its relation to galaxy clustering, an HOD model can tell us much more about a galaxy population, than just the two parameters $M_1$ and $\alpha$. 

For $\lambda(k,M) \rightarrow 1$ (which corresponds to large separations $r$ in real-space) the two halo term simplifies to
\begin{equation}
P_{2h}(k,r) \approx b^2_{\rm eff}P_m(k),
\end{equation}
where the effective bias is the galaxy number weighted halo bias factor~\citep{Tinker:2004gf}
\begin{equation}
b_{\rm eff} = \frac{1}{n_g}\int_0^{\infty}dM\frac{dn(M)}{dM}b_h(M)N_{t}(M).
\label{eq:beff}
\end{equation}
We can also ask what is the average group dark matter halo mass for a specific set of galaxies (often called host-halo mass). Such a quantity can be obtained as
\begin{equation}
M_{\rm eff} = \frac{1}{n_g}\int^{\infty}_0 dM\frac{dn(M)}{dM}MN_{t}(M),
\label{eq:Meff}
\end{equation} 
which represents a weighted sum over the halo mass function with the HOD number as a weight. 

The averaged ratio of satellite galaxies to the total number of galaxies is given by 
\begin{equation}
f_{s} = \frac{\int^{\infty}_0 dM \frac{dn(M)}{dM} N_{s}(M)}{\int^{\infty}_0 dM \frac{dn(M)}{dM} N_{c}(M)\left[1 + N_{s}(M)\right]}
\end{equation}
and the central galaxy ratio is given by $f_c = 1-f_s$.

\section{HOD parameter fits}
\label{sec:hodfits}

\begin{table*}
\begin{center}
\caption{Summary of the best fitting parameters for HOD fits to the four 6dFGS sub-samples ($S_1$-$S_4$) with the fitting range $0.1 < r_p < 40h^{-1}\,$Mpc. The last column shows the reduced $\chi^2$ derived from a fit to $24$ bins with $2$ free parameters and indicates good fits for all sub-samples. Error bars on the HOD parameters correspond to $1\sigma$, derived from the marginalised distributions. The sample $S_2'$ uses a different fitting range of $0.1 < r_p < 20h^{-1}\,$Mpc for the largest of our sub-samples, $S_2$. Furthermore we include a special fit to $S_2$ labeled $S_2^{\sigma_8=0.9}$, where we change our standard assumption of $\sigma_8=0.8$ to $\sigma_8=0.9$. The satellite fraction $f_s$, the effective dark matter halo mass $M_{\rm eff}$ and the effective galaxy bias $b_{\rm eff}$ are derived parameters (see section~\ref{sec:derived}). The bias depends on our initial assumption of $\sigma_8$ and hence this parameter should be treated as $b_{\rm eff}\times (\sigma_8/0.8)$.}
\begin{tabular}{cccccccc}
\hline
sample & $\log_{10}\left(\frac{M_1}{h^{-1}M_{\odot}}\right)$ & $\alpha$ & $\log_{10}\left(\frac{M_{\rm min}}{h^{-1}M_{\odot}}\right)$ & $f_{s}$ & $\log_{10}\left(\frac{M_{\rm eff}}{h^{-1}M_{\odot}}\right)$ & $b_{\rm eff}$ & $\chi^2/\rm d.o.f.$\\
\hline
$S_1$ & $13.396\pm0.017$ & $1.214\pm0.031$ & $12.0478\pm0.0049$ & $0.2106\pm0.0078$ & $13.501\pm0.015$ & $1.2704\pm0.0087$ & $13.5/(24-2) = 0.61$\\
$S_2$ & $13.568\pm0.013$ & $1.270\pm0.027$ & $12.2293\pm0.0043$ & $0.1879\pm0.0062$ & $13.532\pm0.012$ & $1.3443\pm0.0063$ & $16.6/(24-2) = 0.75$\\
$S_3$ & $13.788\pm0.012$ & $1.280\pm0.029$ & $12.4440\pm0.0039$ & $0.1578\pm0.0061$ & $13.546\pm0.012$ & $1.4002\pm0.0062$ & $21.4/(24-2) = 0.97$\\
$S_4$ & $14.022\pm0.011$ & $1.396\pm0.033$ & $12.6753\pm0.0032$ & $0.1243\pm0.0058$ & $13.617\pm0.012$ & $1.5170\pm0.0077$ & $24.4/(24-2) = 1.11$\\
\hline
$S_2^{\sigma_8=0.9}$ & $13.605\pm0.013$ & $1.193\pm0.023$ & $12.2382\pm0.0048$ & $0.2034\pm0.0077$ & $13.649\pm0.014$ & $1.2151\pm0.0065$ & $19.6/(24-2) = 0.89$\\
$S_2'$ & $13.568\pm0.013$ & $1.282\pm0.027$ & $12.2291\pm0.0042$ & $0.1890\pm0.0074$ & $13.542\pm0.010$ & $1.3492\pm0.0065$ & $19.1/(21-2) = 1.01$\\
\hline
\hline
\end{tabular}
\label{tab:results}
\end{center}
\end{table*}

To compare the HOD models to the data we use the same fitting range as for the power law fits earlier ($0.1 < r_p < 40\,h^{-1}\,$Mpc). The free parameters of the fit are $M_1$ and $\alpha$ as described in the last section and hence we have the same number of free parameters as for the power law fits.

\begin{figure}
\begin{center}
\vspace{0.45cm}
\epsfig{file=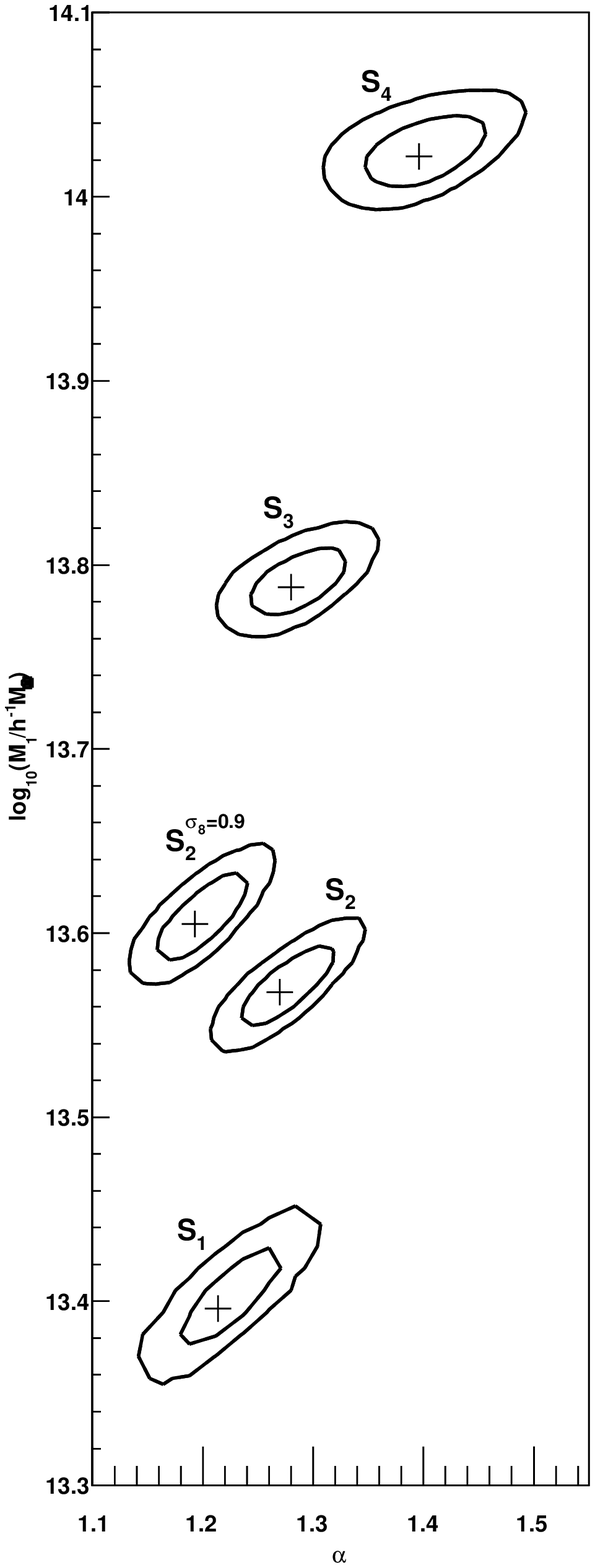,width=8.03cm}
\vspace{0.cm}
\caption{Two dimensional probability distribution for $\log_{10}(M_1)$ and $\alpha$ for the HOD fits to the four 6dFGS volume-limited sub-sample ($S_1$-$S_4$). We include two fits for the largest sample $S_2$, where for $S_2^{\sigma_8=0.9}$ we change our standard assumption of $\sigma_8=0.8$ to $\sigma_8=0.9$. The parameters derived from the fits are summarised in Table~\ref{tab:results}. The best fitting values are marker with black crosses.}
\label{fig:chi2_alpha_M0}
\end{center}
\end{figure}

All the fitting results are summarised in Table~\ref{tab:results} and Figure~\ref{fig:xi_hod_all}. The reduced $\chi^2$ in the last column of Table~\ref{tab:results} indicates a good fit to the data in all cases and the ratio of data to best fitting power law in the lower panels of Figure~\ref{fig:xi_hod_all} shows that the HOD model reproduces the double peak structure present in the data. We also note that the reduced $\chi^2$ in case of the HOD fits is uniformly lower than for the power law fits indicating a better fit to the data for all sub-samples. 

Figure~\ref{fig:chi2_alpha_M0} shows the 2D probability distributions in $M_1$ and $\alpha$ for the four different volume-limited sub-samples. There is a strong trend of increasing $M_1$ with stellar mass and a weaker but still significant trend of increasing $\alpha$ with increasing stellar mass. This indicates that dark matter halos that host a central galaxy with higher stellar mass have their first satellite on average at a larger dark matter halo mass. However the number of satellites increases more steeply with halo mass for samples with higher stellar mass. This trend indicates that halos of higher mass have greater relative efficiency at producing multiple satellites. Similar trends were found in sub-samples of SDSS galaxies by~\citealt{Zehavi:2004ii}. As a test for the sensitivity of the fitting results to the upper fitting limit we performed a fit to sub-sample $S_2$ with the fitting range $0.1 < r_p < 20h^{-1}\,$Mpc. All parameters agree within $1\sigma$ with the fit to the larger fitting range. We called this fit $S_2'$ and included it in Table~\ref{tab:results}.

\begin{figure}
\begin{center}
\epsfig{file=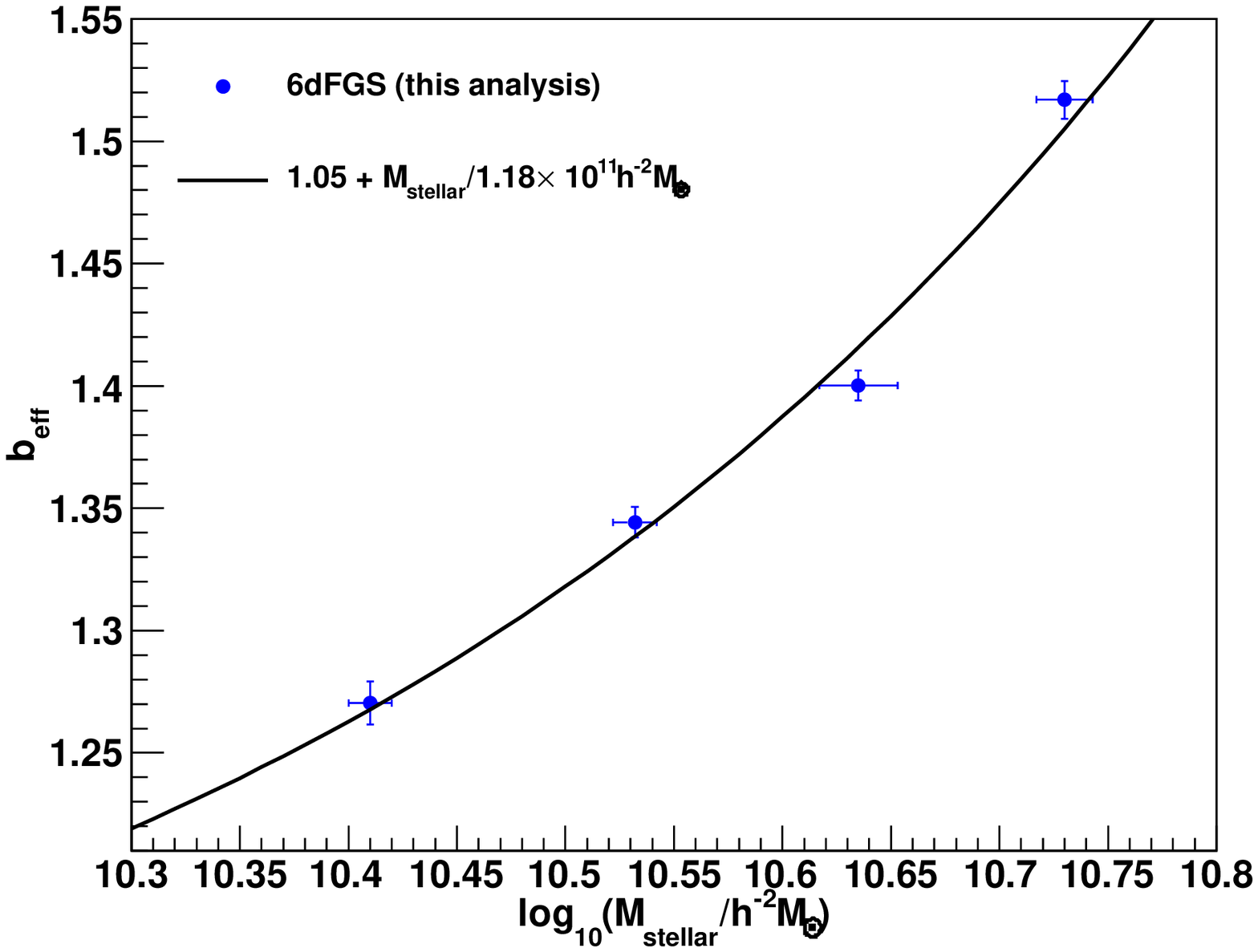,width=8cm}
\caption{The effective galaxy bias (see eq.~\ref{eq:beff}) as a function of log stellar mass for 6dFGS (blue data points). The increase in galaxy bias with stellar mass can be described by eq.~\ref{eq:bias2} which corresponds to the black line.}
\label{fig:beff}
\end{center}
\end{figure}

Table~\ref{tab:results} also includes derived parameters like the minimum dark matter halo mass, $M_{\rm min}$, the satellite fraction, $f_s$, the effective dark matter halo mass, $M_{\rm eff}$ and the effective galaxy bias, $b_{\rm eff}$. The errors on these parameters are calculated as the $68\%$-confidence level of their 1D probability distribution. 
The effective dark matter halo mass seems to be almost constant for all samples while the minimum dark matter halo mass $M_{\rm min}$ increases with stellar mass. The satellite fraction $f_{\rm s}$ decreases with increasing stellar mass, indicating that galaxies with high stellar mass have a higher probability to be central galaxies. The increasing effective galaxy bias indicates that galaxies with higher stellar mass are more strongly clustered and hence reside in high density regions of the Universe. The increasing effective galaxy bias is well described by the following form:
\begin{equation}
b(M_{\rm stellar}) = \left(1.05 + M_{\rm stellar}/M_*\right)\times(0.8/\sigma_8)
\label{eq:bias2}
\end{equation} 
with $M_* = 1.18\times 10^{11}h^{-2}M_{\odot}$. We compare this function to the measurements in Figure~\ref{fig:beff}. The absolute stellar masses are subject to significant uncertainties and different methods to derive stellar masses can come to very different conclusions. However, most stellar mass estimates are related by a simple constant offset, and relation~\ref{eq:bias2} can be scaled accordingly. The equation above is only valid for the stellar mass range probed in this analysis ($M_{\rm stellar} = 2.6 - 5.4\times10^{10}h^{-2}\,M_{\odot}$) since the underlying dynamics are most likely not captured in eq.~\ref{eq:bias2}.

We will discuss the implications of all these results in the next sections. First we will derive 6dFGS mock samples using different semi-analytic models. We will then compare the predictions from these semi-analytic models to our data followed by a comparison to other studies.

\section{semi-analytic mock catalogues}
\label{sec:sam}

To compare our result with theory, we derive 6dFGS mock catalogues from two different semi-analytical models~\citep{Croton:2005fe,Bower:2005vb}, both based on the Millennium Simulation~\citep{Springel:2005nw} publicly available through the Millennium Simulation database\footnote{http://gavo.mpa-garching.mpg.de/MyMillennium/\\http://galaxy-catalogue.dur.ac.uk:8080/Millennium/}. Semi-analytic models are based on an underlying N-body simulation together with theoretically and observationally motivated descriptions of gas cooling, star formation and feedback processes.

The Millennium Simulation is a dark matter only N-body simulation which traces the hierarchical evolution of $2160^3$ particles in a periodic box of $500^3 h^{-3}\,$Mpc$^3$ from redshift $z = 127$ to $z = 0$. The underlying cosmological model follows WMAP1 cosmology~\citep{Spergel:2003cb} given by a matter density of $\Omega_m = \Omega_{dm} + \Omega_b = 0.25$, a cosmological constant of $\Omega_{\Lambda} = 0.75$, a Hubble constant of $H_0=75\,$km/s/Mpc, a spectral index of $n_s = 1$ and a r.m.s. of matter fluctuations in $8h^{-1}\,$Mpc spheres of $\sigma_8 = 0.9$. 
The individual particle mass of the simulation is $8.6\times 10^8 h^{-1}M_{\odot}$ and halos and sub-halos are identified from the spatial distribution of dark matter particles using a standard friends-of-friends algorithm and the SUBFIND algorithm~\citep{Springel:2000qu}. All sub-halos are then linked together to construct the halo merger trees which represent the basic input of the semi-analytic models.

Here we are using the $z=0$ output of the Millennium Simulation. To ensure that the stellar masses are calculated in a consistent and comparable way, we implement the following procedure to derive the 6dFGS mock catalogues:
\begin{enumerate}
\item We apply the 6dFGS $K$-band apparent magnitude limits of $8.85 \leq K \leq 12.75$ to the full $500h^{-3}\,$Mpc$^3$ simulation box. We have to use the $K$-band instead of the $J$-band, which is actually used in this analysis, because none of the semi-analytic models provide $J$-band magnitudes. However, the 6dFGS $J$-band and $K$-band samples have significant overlap, meaning that almost all galaxies which have a $K$-band magnitude also have a $J$-band magnitude. We also account for sky- and magnitude incompleteness.
\item We re-calculate stellar masses for each galaxy using the technique described in section~\ref{sec:stellar} but with the corresponding $K$-band relations~\citep{Bell:2000jt}, instead of $J$-band. The re-calculation of the stellar masses ensures that potential disagreement with our measurement is not caused by a different technique of deriving stellar masses or a different assumptions about the IMF. 
\item We apply the same redshift and stellar mass limits to the semi-analytic catalogues, which we used to produce the four volume-limited samples in the 6dFGS dataset (see Table~\ref{tab:vls}).
\end{enumerate}
We found that all mock 6dFGS catalogues derived from these semi-analytic models contain fewer galaxies than the data sample (by about $40\%$). This could be related to the slightly different cosmology used in these simulations, which should have its largest impact on large clusters, which are sampled in 6dFGS. We are not attempting to correct for such differences in the cosmological model. The aim of this part of our analysis is to test the current predictive power of semi-analytic models.

Semi-analytic models are often grouped into $``$Durham models$"$ and $``$Munich models$"$. The~\citet{Bower:2005vb} model belongs in the group of $"$Durham models$"$ while the~\citet{Croton:2005fe} model belongs in the group of $"$Munich models$"$.

\subsection{Durham models}

In the Durham models, merger trees are produced following~\citet{Helly:2003} which are independent of those generated by~\citet{Springel:2005nw}. When the satellite galaxy falls below a certain distance to the central galaxy given by $R_{\rm merge} = r_{c} + r_s$, where $r_c$ and $r_s$ are the half mass radii of the central and satellite galaxy, respectively, the satellite and central galaxy are treated as one. The largest of the galaxies contained within this new combined dark matter halo is assumed to be the central galaxy~\citep{Benson:2001au}, whilst all other galaxies within the halo are satellites. The dynamical friction and tidal stripping which are present in such a system are modelled analytically. These models are based on NFW density profiles for the central halo as well as the satellite halo, while galaxies are modelled as a disc plus spheroid. 

\subsection{Munich models}

The Munich models are based on the original merger trees by~\citet{Springel:2005nw}. One of the key differences between these merger trees and the ones used in the Durham models is that the Munich models explicitly follow dark matter halos even after they are accreted onto larger systems, allowing the dynamics of satellite galaxies residing in the in-falling halos to be followed until the dark matter substructure is destroyed. The galaxy is than assigned to the most bound particle of the sub-halo at the last time the sub-halo could be identified. 

In the Munich models, a two-mode formalism is adopted for active galactic nuclei (AGNs), wherein a high-energy, or $"$quasar$"$ mode occurs subsequent to mergers, and a constant low-energy $"$radio$"$ mode suppresses cooling flows due to the interaction between the gas and the central black hole~\citep{Croton:2005fe}. In the quasar model, accretion of gas onto the black hole peaks at $z\sim 3$, while the radio mode reaches a plateau at $z\sim 2$. AGN feedback is assumed to be efficient only in massive halos, with supernova feedback being more dominant in lower-mass halos.

The galaxies in the~\citet{Croton:2005fe} model can be of three different types: central galaxies (type $0$), satellites of type $1$ and satellites of type $2$. Satellites of type $1$ are associated with dark matter substructures, which usually refers to recently merged halos. Satellites of type $2$ are instead galaxies whose dark matter halo has completely merged with a bigger halo and are not associated with a substructure. We treat both type $2$ and type $1$ as satellite galaxies.

\subsection{Testing semi-analytic models}

In their original paper~\citet{Bower:2005vb} compare the K-band luminosity function, galaxy stellar mass function, and cosmic star formation rate with high-redshift observations. They find that their model matches the observed mass and luminosity functions reasonably well up to $z \approx 1$. \citet{Kitzbichler:2006ec} compare the magnitude counts in the $b_J$, $r_F$, $I$ and $K$-bands, redshift distributions for $K$-band selected samples, $b_J$- and $K$-band luminosity functions, and galaxy stellar mass function from~\citet{Croton:2005fe} and the very similar model by~\citet{De Lucia:2006vua} with high redshift measurements. They find that the agreement of these models with high-redshift observations is slightly worse than that found for the Durham models. In particular, they find that the Munich models tend to systematically overestimate the abundance of relatively massive galaxies at high redshift. 

\citet{Snaith:2011fy} compared four different semi-analytic models~\citep{De Lucia:2006vua,Bower:2005vb,Bertone:2007sj,Font:2008pc}, with observations and found that all models show a shallower, wider magnitude gap, between the brightest group galaxy and the second brightest, compared to observations.

\citet{delaTorre:2010nm} compared measurements of VVDS with the model by~\citet{De Lucia:2006vua}. They found that the model reproduced the galaxy clustering at $z > 0.8$ as well as the magnitude counts in most bands. However the model failed in reproducing the clustering strength of red galaxies and the $b_J-I$ colour distribution. The model tends to produce too many relatively bright red satellites galaxies, a fact that has been reported in other studies as well (over-quenching problem:~\citealt{Weinmann:2006cq,Kimm:2008rp,Liu:2009rm}).

\section{Discussion}
\label{sec:results}

\subsection{Effective halo mass and satellite fraction}
\label{sec:Meff}

As is evident from Table~\ref{tab:results}, the 6dFGS galaxies sit in massive central dark matter halos and most of our galaxies are central galaxies in these halos with only a small fraction being satellite galaxies. Because of the way 6dFGS galaxies are selected, the large majority are red elliptical galaxies and hence our findings agree very well with previous studies, which also found that such galaxies are strongly clustered and therefore must reside in high density regions.

\begin{figure}
\begin{center}
\epsfig{file=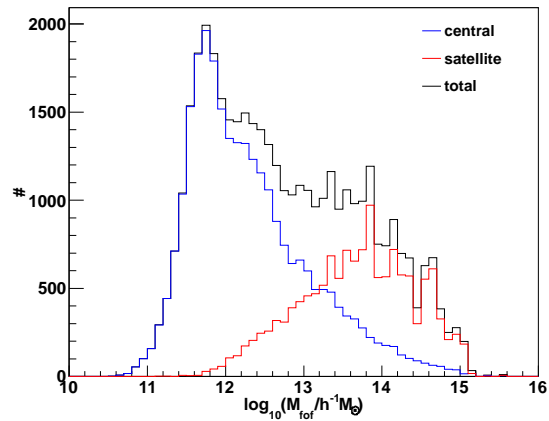,width=8cm}
\caption{This plot shows the friends-of-friends halo mass $\log_{10}( M_{\rm fof})$ of the satellite and central galaxies from the~\citet{Bower:2005vb} semi-analytic catalogue with the 6dFGS selection. The bimodal distribution is caused by a dominant fraction of halos, in which the 6dFGS galaxies are the central galaxy and a smaller fraction with larger halo mass, in which the 6dFGS galaxies are satellite galaxies.}
\label{fig:Meff_mock}
\end{center}
\end{figure}

First we will discuss the effective dark matter halo mass, which represents the effective group or cluster mass for the distribution of galaxies (not the mass of the individual halos which host the galaxies). This quantity appears almost constant for our four different sub-samples. The increasing central galaxy mass ($M_{\rm min}$) with stellar mass is offset by the decreasing satellite fraction, resulting in a fairly constant $M_{\rm eff}$. While the HOD model allows us to derive only averaged parameters for each sample, semi-analytic models directly connect dark matter halo masses with single galaxies. Figure~\ref{fig:Meff_mock} shows the distribution of central and satellite galaxies as a function of friends-of-friends (fof) halo mass derived from the~\citet{Bower:2005vb} semi-analytic catalogue together with the 6dFGS selection criteria. The catalogue contains fewer satellites than central, but the satellites sit in very massive dark matter halos and hence have a significant impact on $M_{\rm eff}$. This plot shows that most 6dFGS galaxies sit in $10^{11}-10^{12}h^{-1}\,M_{\odot}$ halos, while the satellites sit in very massive groups and clusters of up to $10^{15}h^{-1}\,M_{\odot}$. While the median would be around $10^{12}h^{-1}\,M_{\odot}$, the effective mass $M_{\rm eff}$ from the HOD model is the mean of this distribution, which is pushed to very large values by the satellite fraction.

\begin{table}
\begin{center}
\caption{Summary of parameters derived from different semi-analytic models. We impose the 6dFGS $K$-band apparent magnitude selection ($J$-band is not available for these semi-analytic models) as well as correct for incompleteness. The stellar masses are re-calculated using the technique described in section~\ref{sec:stellar}. To calculate the effective dark matter halo mass we used the friends-of-friends halo mass. The stellar mass and the effective halo mass are calculated as the mean of the distribution.}
\begin{tabular}{cccc}
\hline
sample & $f_s$ & $\log_{10}\left(\frac{\langle M_{\rm stellar}\rangle}{h^{-2}M_{\odot}}\right)$ & $\log_{10}\left(\frac{M_{\rm eff}}{h^{-1}M_{\odot}}\right)$\\
\hline
\citep{Croton:2005fe}\\
\hline
S1 & $0.138$ & $10.38$ & $13.400$\\
S2 & $0.099$ & $10.48$ & $13.389$\\
S3 & $0.085$ & $10.57$ & $13.507$\\
S4 & $0.090$ & $10.67$ & $14.013$\\
\hline
\citep{Bower:2005vb}\\
\hline
S1 & $0.28$ & $10.36$ & $13.695$\\
S2 & $0.25$ & $10.47$ & $13.661$\\
S3 & $0.22$ & $10.58$ & $13.674$\\
S4 & $0.18$ & $10.68$ & $13.777$\\
\hline
\hline
\end{tabular}
\label{tab:sams}
\end{center}
\end{table}

\begin{figure}
\begin{center}
\epsfig{file=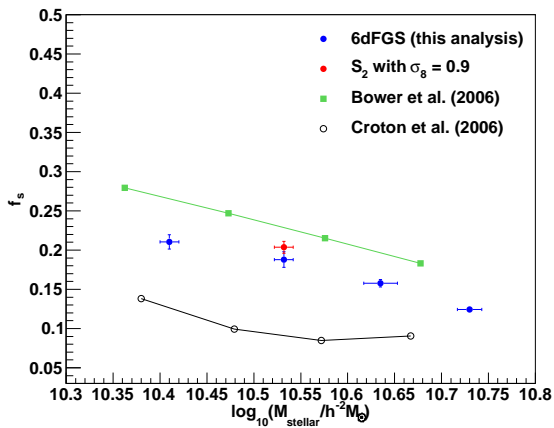,width=8cm}
\caption{Satellite fraction of the 6dFGS stellar mass sub-samples (blue) and the corresponding semi-analytic sub-samples based on the Millennium simulation~\citep{Croton:2005fe, Bower:2005vb}. The red data point represents a fit to sample $S_2$ which assumes $\sigma_8=0.9$, while the blue data points assume $\sigma_8=0.8$.}
\label{fig:cfsat}
\end{center}
\end{figure}

We calculated the effective halo mass $M_{\rm eff} = \langle M_{\rm fof}\rangle$ for the four volume-limited sub-samples derived from the two semi-analytic models and summarise these results in Table~\ref{tab:sams}. While the effective halo mass appears constant in case of the~\citet{Bower:2005vb} model, the~\citet{Croton:2005fe} model shows an increase with increasing stellar mass. These different behaviours for the two different semi-analytic models are most likely connected to the different trend in the satellite fraction shown in Figure~\ref{fig:cfsat}. While the~\citet{Bower:2005vb} model shows a constant decrease in the satellite fraction, the~\citet{Croton:2005fe} model reaches a constant at large stellar mass which causes the effective halo mass to rise. A lower satellite fraction at a fixed stellar mass means that at fixed halo mass the satellite galaxies are less massive~\citep{Weinmann:2006cq,Kimm:2008rp,Liu:2009rm}.

From Figure~\ref{fig:cfsat} we can see that both the~\citet{Croton:2005fe} and the~\citet{Bower:2005vb} model give slightly smaller stellar masses than measured in 6dFGS. This indicates that the semi-analytic catalogues contain too few bright galaxies at low redshift.

Figure~\ref{fig:cfsat} also includes the fit to sample $S_2$ which assumes $\sigma_8=0.9$ (red data point), which agrees with the assumptions of the Millennium Simulation, while the blue data points assume $\sigma_8=0.8$. A larger $\sigma_8$ causes a larger satellite fraction which than causes a larger effective halo mass. This effect seem to bring our results closer to the~\citet{Bower:2005vb} model.

\subsection{The $M_1 - M_{\rm min}$ scaling relation}

\begin{figure*}
\begin{center}
\epsfig{file=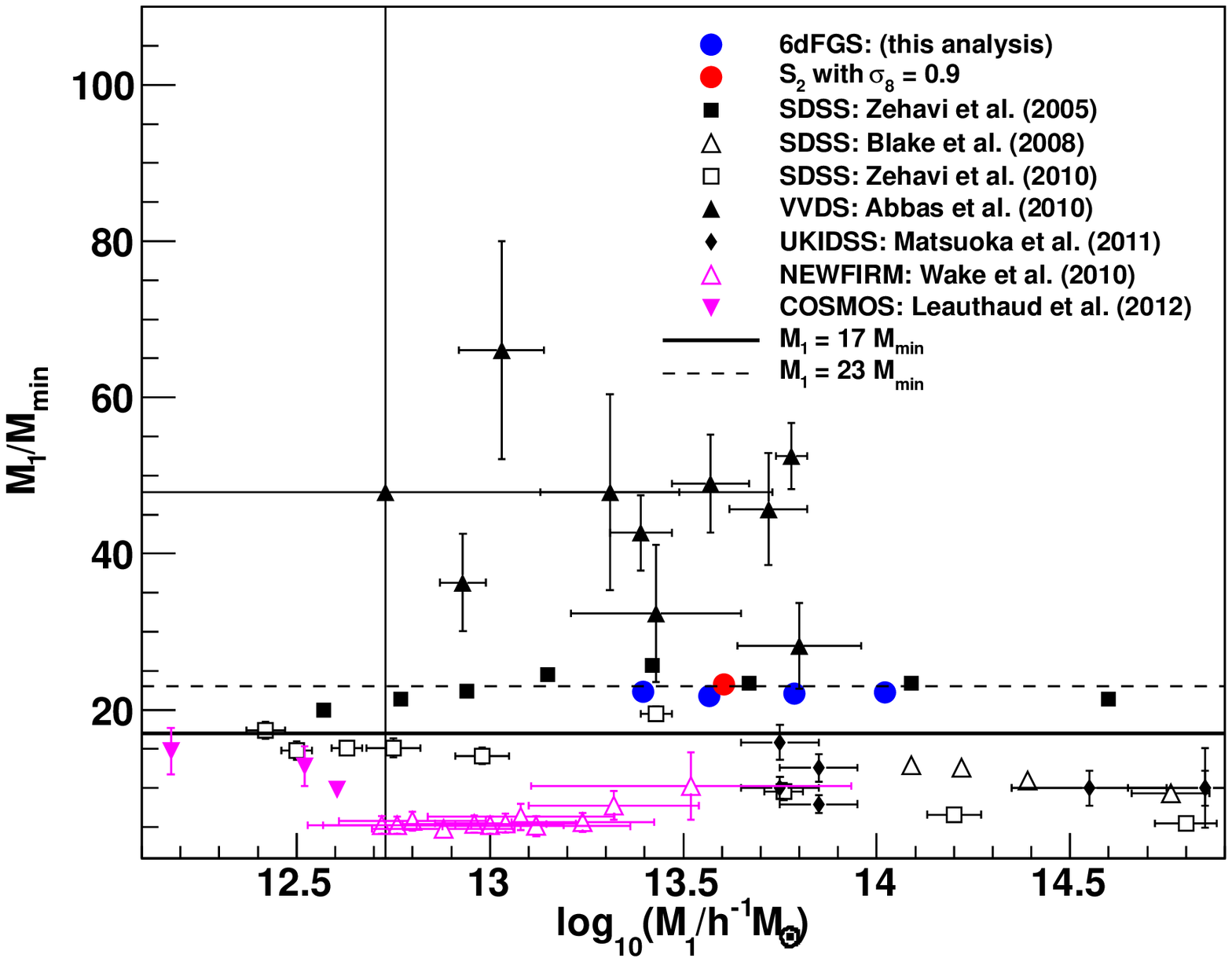,width=14cm}
\caption{The relation between $M_1$ and $M_{\rm min}$ for the four different volume-limited 6dFGS samples (blue data points), compared to~\citet{Zehavi:2004ii} (black solid squares),~\citet{Blake:2007xp} (black open triangles),~\citet{Zehavi:2010bh} (black open squares),~\citet{Abbas:2010hr} (black solid triangles),~\citet{Matsuoka:2010ba} (black open points),~\citet{Wake:2010um} (magenta open triangles) and \citet{Leauthaud:2011rj} (magenta solid triangles). \citet{Zehavi:2004ii} don't report errors on their parameters and the 6dFGS errors are smaller than the data points. All studies which investigate HOD as a function of stellar mass are coloured (including our analysis), while luminosity defined studies are in black.}
\label{fig:M0_Mmin2}
\end{center}
\end{figure*}

The HOD analysis of early data from SDSS by~\citet{Zehavi:2004ii} showed the relation, $M_1 \approx 23\,M_{\rm min}$, between the mass of halos that on average host one additional satellite galaxy, $M_1$ and the minimum dark matter halo mass to host a central galaxy, $M_{\rm min}$. This has been confirmed in subsequent studies ($M_1 \approx 18\,M_{\rm min}$ in~\citealt{Zheng:2007zg} and $M_1 \approx 17\,M_{\rm min}$~\citealt{Zehavi:2010bh}). This relation implies that on average a halo hosting two galaxies of the type studied in their analysis has a mass $\approx 23$ times the mass of a halo hosting only one galaxy of the same type.

\citet{Zehavi:2010bh} also found that this scaling factor is somewhat smaller at the high luminosity end, corresponding to massive halos that host rich groups or clusters. This latter trend likely reflects the relatively late formation of these massive halos, which leaves less time for satellites to merge onto central galaxies and thus lowers the satellite threshold $M_1$.

Theoretical studies of HODs in dark matter simulations~\citep{Kravtsov:2003sg} and those predicted by Smoothed-particle hydrodynamics (SPH) and semi-analytic galaxy formation models~\citep{Zheng:2004id} reveal a similar relation with a scaling factor of $\approx 20$. The large gap between $M_1$ and $M_{\rm min}$ arises because in the low occupation regime, a more massive halo tends to host a more massive central galaxy, rather than multiple smaller galaxies~\citep{Berlind:2002rn}.

\citet{Abbas:2010hr} did a similar study using data from the $I$-band selected VIMOS-VLT Deep Survey (VVDS) and found the ratio $M_1/M_{\rm min}$ to be $\approx 40-50$. The galaxies in this study are at much higher redshift ($z \approx 0.83$) compared to the SDSS galaxies. This result means that in order to begin hosting satellite galaxies, halos sampled by the VVDS survey need to accrete a larger amount of mass compared to SDSS halos. 

\citet{Wake:2010um} also studied high redshift galaxies ($1.1 < z < 1.9$) in the NEWFIRM Medium Band Survey (NMBS) and found the ratio $M_1/M_{\rm min}$ to be $\approx 4-10$. This, together with the result by~\citet{Abbas:2010hr} shows, that this relation is not as fundamental as originally thought, but strongly dependent on the type of halos probed in each analysis.

\citet{Leauthaud:2011rj} studied galaxy clustering in the COSMOS survey using threshold stellar mass samples. Using the equations discussed in~\citep{Leauthaud:2011zt} we can derive $M_1$ and $M_{\rm min}$ from their HOD parametrisation, which is included in Figure~\ref{fig:M0_Mmin2}.

\citet{Matsuoka:2010ba} analysed $\sim60\,000$ massive ($\log_{10}(M_{\rm stellar}/h^{-2}M_{\odot}) > 10.7$) galaxies from the UKIRT Infrared Deep Sky Survey (UKIDSS) and the SDSS II Supernova Survey. This analysis shows a very different clustering amplitude depending on whether the observed and theoretical number densities are matched up or not. This dependency makes a comparison of our results with their derived parameters very difficult. Nevertheless, the ratio $M_1/M_{\rm min}$ in their analysis does not depend significantly on their initial assumptions and hence we included their results in Figure~\ref{fig:M0_Mmin2}\footnote{We use their results, in which they match the observed and theoretical number densities, since this agrees with our method.}. Their results indicate a lower $M_1/M_{\rm min}$ ratio at large $M_1$ consistent with~\citet{Zehavi:2010bh} and~\citet{Blake:2007xp}.


In our analysis we found a scaling relation of $M_1 \approx 22\,M_{\rm min}$ (the exact values are $22.29\pm0.39, 21.81\pm0.30, 22.08\pm0.28$ and $22.22\pm0.25$ for $S_1$-$S_4$, respectively). In Figure~\ref{fig:M0_Mmin2} we compare our results with other stellar mass selected samples (coloured data points) and with luminosity threshold selected samples (black data points). The ratio $M_1/M_{\rm min}$ found in 6dFGS is in agreement with~\citet{Zehavi:2004ii} and slightly larger than~\citet{Zehavi:2010bh}. While~\citet{Zehavi:2004ii} and~\citet{Zehavi:2010bh} study the same type of galaxies,~\citet{Zehavi:2004ii} uses an HOD parameterisation very similar to ours, while~\citet{Zehavi:2010bh} uses a parameterisation based on $5$ free parameters, suggesting that the differences might be related to the parameterisation. 

\subsection{Comparison to~\citet{Mandelbaum:2005nx}}

An alternative method for probing the connection between stellar mass and halo mass is galaxy-galaxy weak lensing. Here we will compare our findings to~\citet{Mandelbaum:2005nx}, who used weak lensing of $\sim 350\,000$ galaxies from SDSS and looked at the dependence of the amplitude of the lensing signature as a function of galaxy type and stellar mass. They used stellar masses derived from the $z$-band magnitude and the ratio $M/L_z$ from~\citet{Kauffmann:2002pn} with the assumption of a~\citet{Kroupa:2000iv} IMF. 

The HOD model employed in there analysis is given by
\begin{equation}
N_t(M) = N_c(M) + N_s(M)
\end{equation}
with the central galaxy number given by eq.~\ref{eq:hod} while the satellite galaxies number is modelled by a step-like function
\begin{equation}
N_s(M) = \begin{cases}	kM & \text{ if }\;M\geq 3M_{\rm min}\\
				    	\frac{kM^2}{3M_{\rm min}} & \text{ if }\;M_{\rm min} \leq M < 3M_{\rm min}\\
					0 & \text{ if }\;M < M_{\rm min}.
	        \end{cases}
	        \label{eq:mandel1}
\end{equation}
The normalisation constant $k$ can be determined by matching the measured satellite fractions $f_s^{\rm ma}$ and is given in appendix~\ref{ap:mandel}. Here we have adjusted the nomenclature to the one used in our analysis. We also note that~\citet{Mandelbaum:2005nx} assumed $\sigma_8 = 0.9$, while we assumed $\sigma_8 = 0.8$ for most of our fits.

From the equation above we can derive the effective halo mass using eq.~\ref{eq:Meff} together with the dark matter halo mass function.

\begin{figure}
\begin{center}
\epsfig{file=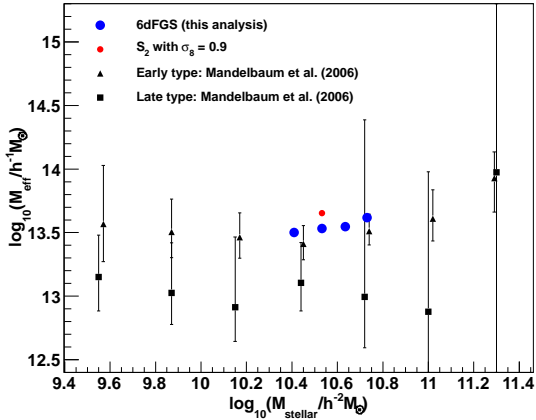,width=8cm}
\caption{The effective dark matter halo mass of the four 6dFGS sub-samples (blue data points) compared to early type (black triangles) and late type (black squares) galaxies from~\citet{Mandelbaum:2005nx}. In red we also include the fit to sample $S_2$ which assumes $\sigma_8=0.9$ in agreement with the assumptions of~\citet{Mandelbaum:2005nx}, while the blue data points assume $\sigma_8=0.8$.}
\label{fig:cMeff_sm}
\end{center}
\end{figure}

\begin{figure}
\begin{center}
\epsfig{file=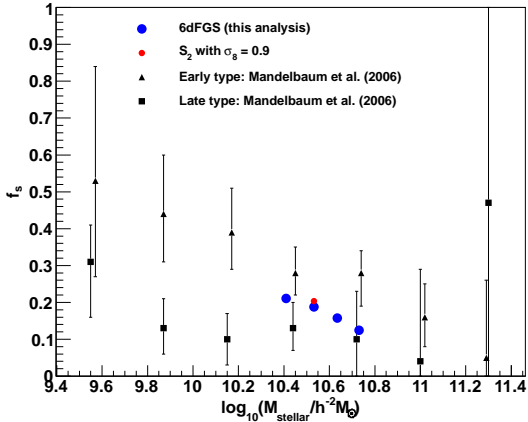,width=8cm}
\caption{The satellite fraction of the four 6dFGS sub-samples (blue data points) compared to early type (black triangles) and late type (black squares) galaxies from~\citet{Mandelbaum:2005nx}. In red we also include the fit to sample $S_2$ which assumes $\sigma_8=0.9$ in agreement with the assumptions of~\citet{Mandelbaum:2005nx}, while the blue data points assume $\sigma_8=0.8$.}
\label{fig:calpha_sm}
\end{center}
\end{figure}

We compare the reported satellite fraction, $f_s^{\rm ma}$, of~\citet{Mandelbaum:2005nx} and the derived effective halo mass, $M_{\rm eff}$, with our results in Figure~\ref{fig:cMeff_sm} and Figure~\ref{fig:calpha_sm}. Figure~\ref{fig:cMeff_sm} shows that the galaxies sampled by~\citet{Mandelbaum:2005nx} have a very similar effective halo mass compared to 6dFGS galaxies. The 6dFGS measurements follow the trend of early-type galaxies in these measurements, while the late types have a slightly smaller effective halo mass. However, the errors in case of the results by~\citet{Mandelbaum:2005nx} don't allow one to distinguish between the early and late type galaxies. Since effective halo mass and satellite fraction are strongly linked, we also compare the satellite fraction of~\citet{Mandelbaum:2005nx} with our results in Figure~\ref{fig:calpha_sm}. Both figures also include the fit to sample $S_2$ where we assumed $\sigma_8=0.9$, instead of the standard $\sigma_8=0.8$, since this agrees with the assumption in~\citet{Mandelbaum:2005nx}. A larger $\sigma_8$ increases the effective halo mass by a small amount and hence does not impact this comparison significantly.

Overall we see very good agreement between our results and the results of~\citet{Mandelbaum:2005nx}, which is reassuring since the two techniques are subject to different systematic uncertainties. We can also emphasise that the 6dFGS results are extremely precise compared to the lensing results.

\subsection{Comparison to other studies}

\citet{Meneux:2007jr} study galaxy cluster mass dependence on stellar mass in the VIMOS-VLT Deep Survey (VVDS) at redshift $0.5 < z < 1.2$. The stellar masses in this sample cover the range $\log_{10}(M_{\rm stellar}/h^{-2}M_{\odot}) = 8.7-10.7$. To quantify the clustering, they use power laws fits to the projected correlation function $w_p(r_p)$ and found an evolution in $r_0$ from $2.76\pm0.17 h^{-1}\,$Mpc at $\log_{10}(M_{\rm stellar}/h^{-2}M_{\odot}) > 8.7$ to $r_0 = 4.28\pm0.45h^{-1}\,$Mpc at $\log_{10}(M_{\rm stellar}/h^{-2}M_{\odot}) > 10.2$. The slope changes over the same range from $\gamma = 1.67\pm 0.08$ to $\gamma = 2.28\pm 0.28$. Comparing the results by~\citet{Meneux:2007jr} with~\citet{Li:2005er} who used the clustering of $200\,000$ SDSS galaxies at $z = 0.15$, showed that the evolution of the amplitude and shape of the correlation function $w_p(r_p)$ with redshift is faster for low stellar mass objects than for high stellar mass objects. At low stellar mass, the amplitude of $w_p(r_p)$ increases by a factor $\sim2-3$ from high to low redshift, while at the high stellar mass range $\log_{10}(M_{\rm stellar}/h^{-2}M_{\odot}) = 10.2-10.7$, the amplitude at $z \sim 0.85$ and $z \sim 0.15$ are very similar, within the error bars. 
In 6dFGS we found a significantly larger clustering amplitude, ranging from $r_0 = 5.14\pm0.23h^{-1}$Mpc at $\log_{10}(M_{\rm stellar}/h^{-2}M_{\odot}) = 10.41$ to $r_0 = 6.21\pm0.17h^{-1}$Mpc at $\log_{10}(M_{\rm stellar}/h^{-2}M_{\odot}) = 10.73$, while our power law index $\gamma$ stays constant at $\gamma \approx 1.84$. This shows that the trend observed in~\citet{Meneux:2007jr} and~\citet{Li:2005er} continues to the 6dFGS redshifts, even for high stellar mass galaxies.

\section{Conclusions}
\label{sec:con}

We present in this paper an analysis of the clustering properties of four stellar mass selected volume-limited sub-samples of galaxies from the 6dF Galaxy Survey. The stellar masses are calculated using the $J$-band magnitude and the $b_J-r_F$ colour following the technique by~\citet{Bell:2000jt}. The average log-stellar mass for the different sub-samples ranges from $\log_{10}(M_{\rm stellar}/h^{-2}\,M_{\odot}) = 10.41-10.73$. Our analysis has the following main results:

\begin{itemize}
\item The projected correlation function, $w_p(r_p)$, for the two low stellar mass sub-samples ($S_1$ and $S_2$) can be described by a power law with an acceptable $\chi^2$, while the two high stellar mass sub-samples ($S_3$ and $S_4$) have a reduced $\chi^2$ of $1.48$ and $1.63$, respectively. We also find patterns in the deviations between the best power law fit and the data, which can naturally be explained within the halo model. This is in agreement with theoretical studies~\citep{Watson:2011cz}, predicting that the disagreement of $w_p(r_p)$ with a power law fit should grow with clustering amplitude.
\item We used an HOD parameterisation with three free parameters ($M_{1}$, $\alpha$, $M_{\rm min}$), representing the typical halo mass, satellite power law index and minimum host halo mass, respectively. The minimum halo mass is fixed by the galaxy number density for all our parameter fits. We performed fits to the projected correlation functions of the four different volume-limited sub-samples. We tested alternative HOD parameterisations, but found that our data does not justify more free parameters. $M_{1}$ and $\alpha$ show increasing trends for increasing stellar mass, with $\log_{10}(M_{1}/h^{-1}\,M_{\odot})$ ranging from $13.4 - 14$ and $\alpha$ from $1.21-1.4$. This means that galaxies with larger stellar mass populate larger dark matter halos, which accrete satellites faster with increasing mass, compared to dark matter halos populated by galaxies with lower stellar mass.
\item From the halo model, we can derive averaged parameters for the four samples such as the satellite fraction, effective dark matter halo mass and the effective galaxy bias. We found that the satellite fraction decreases with stellar mass from $21\%$ at $\log_{10}(M_{\rm stellar}/h^{-2}\,M_{\odot}) = 10.41$ to $12\%$ at $\log_{10}(M_{\rm stellar}/h^{-2}\,M_{\odot}) = 10.73$. The effective dark matter halo mass stays constant at $\log_{10}(M_{\rm eff}/h^{-1}M_{\odot}) \approx 13.55$ for all four sub-samples. The effective galaxy bias increases with stellar mass indicating that galaxies with higher stellar mass reside in denser environments. The increase in the effective galaxy bias can be described by $(1.05 + M_{\rm stellar}/M_*)\times(\sigma_8/0.8)$ with $M_* = 1.18\times 10^{11}h^{-2}M_{\odot}$.
\item We use two semi-analytic models, based on the Millennium Simulation (\citealt{Croton:2005fe} and~\citealt{Bower:2005vb}) to derive 6dFGS mock surveys. We compare the results of these semi-analytic models with our measurements. The~\citet{Croton:2005fe} model under-predicts the satellite fraction, while the~\citet{Bower:2005vb} model is in better agreement with our observations, although it slightly over-predicts the satellite fraction. Since the effective dark matter halo mass is strongly linked to the satellite fraction, the~\citet{Bower:2005vb} model prediction of $M_{\rm eff}$ is again in better agreement with our observations. 6dFGS allows a powerful test of semi-analytic models, because of (1) the robust stellar mass estimates and (2) the focus on $``$red satellites$"$, which semi-analytic models struggled to reproduce in the past. Our results can be used as a new constraint on semi-analytic models in the future. 
\item For the four volume-limited samples we find a constant scaling relation between $M_1$ and $M_{\rm min}$ of $M_1 \approx 22\,M_{\rm min}$, in agreement with studies of SDSS galaxies~\citep{Zehavi:2004ii,Zehavi:2010bh}. This indicates that 6dFGS galaxies populate dark matter halos in a similar way to SDSS galaxies. However, we see a wide spectrum of this ratio ranging from $> 40$ in~\citet{Abbas:2010hr} to $< 10$ in~\citep{Matsuoka:2010ba,Wake:2010um} indicating that this ratio is not universal for all types of galaxies.
\item We compare our results with the results of~\citet{Mandelbaum:2005nx} from galaxy-galaxy weak lensing. We find overall good agreement which represents a valuable crosscheck for these two different clustering measurements. Although our analysis depends on slightly more assumptions and covers a smaller range in stellar mass, the 6dFGS results are extremely precise compared to the lensing results.
\end{itemize}

\section*{Acknowledgements}

The authors thank Alex Merson for providing the random mock generator.  We also acknowledge Rachel Mandelbaum, Darren Croton, Danail Obreschkow, Alan Duffy, Martin Meyer and Michael Brown for helpful discussions. Thanks also to the referee for detailed feedback which helped to improved the final version of this paper.

F.B. is supported by the Australian Government through the International Postgraduate Research Scholarship (IPRS) and by scholarships from ICRAR and the AAO.
The 6dF Galaxy Survey was funded in part by an Australian Research Council Discovery--Projects Grant (DP-0208876), administered by the Australian National University. 

The Millennium Run simulation used in this paper was carried out by the Virgo Supercomputing Consortium at the Computing Centre of the Max-Planck Society in Garching. The semi-analytic galaxy catalogue is publicly available at http://www.mpa-garching.mpg.de/galform/agnpaper and http://galaxy-catalogue.dur.ac.uk:8080/Millennium/.

\setlength{\bibhang}{2em}
\setlength{\labelwidth}{0pt}

\appendix

\section{Normalisation constant in M\lowercase{andelbaum et al. 2006}}
\label{ap:mandel}

We can determine the normalisation constant $k$ of eq.~\ref{eq:mandel1} by matching to the satellite fraction $f_s^{\rm ma}$, which in terms of the mass function $dn(M)/dM$ is given by
\begin{align}
f_s^{\rm ma} &= \frac{\int^{\infty}_0 dM\frac{dn(M)}{dM}N_s(M)}{\int^{\infty}_0dM\frac{dn(M)}{dM} \left[N_c(M) + N_s(M)\right]}\\
&=\frac{kH}{ \int_{M_{\rm min}}^{\infty} dM \frac{dn(M)}{dM}  + kH}
\end{align}
with 
\begin{equation}
\begin{split}
H =& \frac{1}{3M_{\rm min}}\int^{3M_{\rm min}}_{M_{\rm min}}dM\frac{dn(M)}{dM}M^2 +\\ 
&\int^{\infty}_{3M_{\rm min}}dM\frac{dn(M)}{dM}M,
\end{split}
\end{equation}
which leads to
\begin{equation}
k = \frac{f_s^{\rm ma}\int_{M_{\rm min}}^{\infty} dM \frac{dn(M)}{dM}}{(1-f_s^{\rm ma})H},
\end{equation}
where $k$ has units of $1/h^{-1}M_{\odot}$ and is constant for a given set of $f_s^{\rm ma}$ and $M_{\rm min}$.

\newpage

\label{lastpage}


\begin{thebibliography}{99}

\bibitem[\protect\citeauthoryear{Abbas et al.}{2010}]{Abbas:2010hr} 
  Abbas~U. {\it et al.},
  arXiv:1003.6129 [astro-ph.CO].
  
\bibitem[\protect\citeauthoryear{Akaike et al.}{1974}]{Akaike:1974} 
  Akaike~H.
   IEEE Transactions on Automatic Control 19 (6): 716Ð723
  
\bibitem[\protect\citeauthoryear{Bell \& De Jong}{2001}]{Bell:2000jt} 
  Bell~E.~F. and de Jong~R.~S.,
  Astrophys.\ J.\  {\bf 550}, 212 (2001)
  [astro-ph/0011493].
  
\bibitem[\protect\citeauthoryear{Benson et al.}{2002}]{Benson:2001au} 
  Benson~A.~J., Lacey~C.~G., Baugh~C.~M., Cole~S. and Frenk~C.~S.,
  Mon.\ Not.\ Roy.\ Astron.\ Soc.\  {\bf 333}, 156 (2002)
  [astro-ph/0108217].
  
\bibitem[\protect\citeauthoryear{Berlind \& Weinberg}{2001}]{Berlind:2001xk}
  Berlind~A.~A. and Weinberg~D.~H.,
  Astrophys.\ J.\  {\bf 575} (2002) 587
  [arXiv:astro-ph/0109001].
  
\bibitem[\protect\citeauthoryear{Berlind et al.}{2003}]{Berlind:2002rn} 
  Berlind~A.~A. {\it et al.},
  Astrophys.\ J.\  {\bf 593}, 1 (2003)
  [astro-ph/0212357].
  
\bibitem[\protect\citeauthoryear{Bertone et al.}{2007}]{Bertone:2007sj} 
  Bertone~S., De Lucia~G. and Thomas~P.~A.,
  Mon.\ Not.\ Roy.\ Astron.\ Soc.\  {\bf 379}, 1143 (2007)
  [astro-ph/0701407].
  
\bibitem[\protect\citeauthoryear{Beutler et al.}{2011}]{Beutler:2011hx}
  Beutler~F. {\it et al.},
  Mon.\ Not.\ Roy.\ Astron.\ Soc.\  {\bf 416}, (2011) 3017B
  arXiv:1106.3366 [astro-ph.CO].
  
\bibitem[\protect\citeauthoryear{Beutler et al.}{2012}]{Beutler:2012px} 
  Beutler~F. {\it et al.},
  Mon.\ Not.\ Roy.\ Astron.\ Soc.\  {\bf 423}, (2012) 3430B
  arXiv:1204.4725 [astro-ph.CO].
  
\bibitem[\protect\citeauthoryear{Blake, Collister \& Lahav}{2008}]{Blake:2007xp} 
  Blake~C., Collister~A. and Lahav~O.,
  Mon.\ Not.\ Roy.\ Astron.\ Soc.\  {\bf 385}, 1257 (2008)
  [arXiv:0704.3377 [astro-ph]].
  
\bibitem[\protect\citeauthoryear{Bower et al.}{2006}]{Bower:2005vb} 
  Bower~R.~G., Benson~A.~J., Malbon~R., Helly~J.~C., Frenk~C.~S., Baugh~C.~M., Cole~S. and Lacey~C.~G.,
  Mon.\ Not.\ Roy.\ Astron.\ Soc.\  {\bf 370}, 645 (2006)
  [astro-ph/0511338].
  
\bibitem[\protect\citeauthoryear{Brown et al.}{2008}]{Brown:2008eb} 
  Brown~M.~J.~I. {\it et al.},
  Astrophys.\ J.\  {\bf 682}, 937 (2008)
  [arXiv:0804.2293 [astro-ph]].
  
\bibitem[\protect\citeauthoryear{Bruzual \& Charlot}{1993}]{Bruzual A.:1993is} 
  Bruzual~G. and Charlot~S.,
  Astrophys.\ J.\  {\bf 405}, 538 (1993).
  
\bibitem[\protect\citeauthoryear{Bullock et al.}{1999}]{Bullock:1999he}
  Bullock~J.~S. {\it et al.},
  Mon.\ Not.\ Roy.\ Astron.\ Soc.\  {\bf 321} (2001) 559
  [arXiv:astro-ph/9908159].
  
\bibitem[\protect\citeauthoryear{Coil et al.}{2006}]{Coil:2005ku}
  Coil~A.~L. {\it et al.},
  Astrophys.\ J.\  {\bf 638} (2006) 668
  [astro-ph/0507647].
  
\bibitem[\protect\citeauthoryear{Conroy, Wechsler \& Kravtsov}{2006}]{Conroy:2005aq} 
  Conroy~C., Wechsler~R.~H. and Kravtsov~A.~V.,
  Astrophys.\ J.\  {\bf 647}, 201 (2006)
  [astro-ph/0512234].
  
\bibitem[\protect\citeauthoryear{Cooray \& Sheth}{2002}]{Cooray:2002dia}
  Cooray~A. and Sheth~R.~K.,
  Phys.\ Rept.\  {\bf 372} (2002) 1
  [arXiv:astro-ph/0206508].
  
\bibitem[\protect\citeauthoryear{Croton et al.}{2006}]{Croton:2005fe} 
  Croton~D.~J.{\it et al.},
  Mon.\ Not.\ Roy.\ Astron.\ Soc.\  {\bf 365}, 11 (2006)
  [astro-ph/0508046].
  
\bibitem[\protect\citeauthoryear{Davis \& Peebles}{1982}]{Davis:1982gc}
  Davis~M. and Peebles~P.~J.~E.,
  Astrophys.\ J.\  {\bf 267} (1982) 465.
  
\bibitem[\protect\citeauthoryear{de la Torre et al.}{2010}]{delaTorre:2010nm} 
  de la Torre~S. {\it et al.},
  arXiv:1010.0360 [astro-ph.CO].
  
\bibitem[\protect\citeauthoryear{De Lucia \& Blaizot}{2007}]{De Lucia:2006vua}
  De Lucia~G. and Blaizot~J.,
  Mon.\ Not.\ Roy.\ Astron.\ Soc.\  {\bf 375} (2007) 2
  [astro-ph/0606519].
  
\bibitem[\protect\citeauthoryear{Drory et al.}{2004}]{Drory:2004ib} 
  Drory~N., Bender~R. and Hopp~U.,
  Astrophys.\ J.\  {\bf 616}, L103 (2004)
  [astro-ph/0410084].
  
\bibitem[\protect\citeauthoryear{Duffy et al.}{2008}]{Duffy:2008pz}
  Duffy~A.~R., Schaye~J., Kay~S.~T. and Dalla Vecchia~C.,
  Mon.\ Not.\ Roy.\ Astron.\ Soc.\  {\bf 390}, L64 (2008)
  [arXiv:0804.2486 [astro-ph]].
  
\bibitem[\protect\citeauthoryear{Font et al.}{2008}]{Font:2008pc}
  Font~A.~S. {\it et al.},
  Mon.\ Not.\ Roy.\ Astron.\ Soc.\  {\bf 389} (2008) 1619
  [arXiv:0807.0001 [astro-ph]].
  
\bibitem[\protect\citeauthoryear{Gallazzi \& Bell}{2009}]{Gallazzi:2009aj} 
  Gallazzi~A. and Bell~E.~F.,
  Astrophys.\ J.\ Suppl.\  {\bf 185}, 253 (2009)
  [arXiv:0910.1591 [astro-ph.CO]].
  
\bibitem[\protect\citeauthoryear{Grillo et al.}{2008}]{Grillo:2007kg} 
  Grillo~C., Gobat~R., Rosati~P. and Lombardi~M.,
  A\&A  {\bf 477}, L25 (2008)
  [arXiv:0712.0680 [astro-ph]].
  
\bibitem[\protect\citeauthoryear{Guo et al.}{2010}]{Guo:2009fn} 
  Guo~Q., White~S., Li~C. and Boylan-Kolchin~M.,
  Mon.\ Not.\ Roy.\ Astron.\ Soc.\  {\bf 404}, 1111 (2010)
  [arXiv:0909.4305 [astro-ph.CO]].
  
\bibitem[\protect\citeauthoryear{Hauser \& Peebles}{1973}]{Hauser:1973} 
Hauser~M.~G., \& Peebles, P.~J.~E.,
ApJ, 185, 757 (1973)

\bibitem[\protect\citeauthoryear{Hauser \& Peebles}{1974}]{Hauser:1974} 
Peebles, P.~J.~E., \& Hauser, M.~G.,
ApJS, 28, 19 (1974)

\bibitem[\protect\citeauthoryear{Hawkins et al.}{2003}]{Hawkins:2002sg} 
  Hawkins~E. {\it et al.},
  Mon.\ Not.\ Roy.\ Astron.\ Soc.\  {\bf 346}, 78 (2003)
  [astro-ph/0212375].

\bibitem[\protect\citeauthoryear{Helly et al.}{2003}]{Helly:2003} 
Helly~J.~C., Cole~S., Frenk~C.~S., et al.\ 2003, 
  Mon.\ Not.\ Roy.\ Astron.\ Soc.\  {\bf 338}, 903 
  
\bibitem[\protect\citeauthoryear{Jarrett et al.}{2000}]{Jarrett:2000me}
  Jarrett~T.~H., Chester~T., Cutri~R., Schneider~S., Skrutskie~M. and Huchra~J.~P.,
  Astron.\ J.\  {\bf 119} (2000) 2498
  [arXiv:astro-ph/0004318].
  
\bibitem[\protect\citeauthoryear{Jing, Mo \& Borner}{1998}]{Jing:1997nb} 
  Jing~Y.~P., Mo~H.~J. and Borner~G.,
  Astrophys.\ J.\  {\bf 494}, 1 (1998)
  [astro-ph/9707106].

\bibitem[\protect\citeauthoryear{Jones et al.}{2004}]{Jones:2004zy}
  Jones~D.~H. {\it et al.},
  MNRAS {\bf 355} (2004) 747
  [arXiv:astro-ph/0403501].
  
\bibitem[\protect\citeauthoryear{Jones et al.}{2005}]{Jones:2005ya} 
  Jones~D.~H., Saunders~W., Read~M. and Colless~M.,
  [astro-ph/0505068].

\bibitem[\protect\citeauthoryear{Jones et al.}{2006}]{Jones:2006xy}
  Jones~D.~H., Peterson~B.~A., Colless~M. and Saunders~W.,
  MNRAS {\bf 369} (2006) 25
  [Erratum-ibid.\  {\bf 370} (2006) 1583]
  [arXiv:astro-ph/0603609].

\bibitem[\protect\citeauthoryear{Jones et al.}{2009}]{Jones:2009yz}
  Jones~D.~H. {\it et al.},
  arXiv:0903.5451 [astro-ph.CO].
  
\bibitem[\protect\citeauthoryear{Kannappan \& Gawiser}{2007}]{Kannappan:2007ys}
  Kannappan~S.~J. and Gawiser~E.,
  Astrophys.\ J.\  {\bf 657} (2007) L5
  [astro-ph/0701749 [ASTRO-PH]].
  
\bibitem[\protect\citeauthoryear{Kauffmann, White \& Guiderdoni}{1993}]{Kauffmann:1993gv} 
  Kauffmann~G., White~S.~D.~M. and Guiderdoni~B.,
  Mon.\ Not.\ Roy.\ Astron.\ Soc.\  {\bf 264}, 201 (1993).
  
\bibitem[\protect\citeauthoryear{Kauffmann et al.}{2003}]{Kauffmann:2002pn}
  Kauffmann~G. {\it et al.}  [SDSS Collaboration],
  Mon.\ Not.\ Roy.\ Astron.\ Soc.\  {\bf 341} (2003) 33
  [astro-ph/0204055].
  
\bibitem[\protect\citeauthoryear{Kimm et al.}{2009}]{Kimm:2008rp}
  Kimm~T. {\it et al.},
  arXiv:0810.2794 [astro-ph].
  
\bibitem[\protect\citeauthoryear{Kitzbichler \& White}{2007}]{Kitzbichler:2006ec} 
  Kitzbichler~M.~G. and White~S.~D.~M.,
  Mon.\ Not.\ Roy.\ Astron.\ Soc.\  {\bf 376}, 2 (2007)
  [astro-ph/0609636].
  
\bibitem[\protect\citeauthoryear{Komatsu et al.}{2011}]{Komatsu:2010fb}
  Komatsu~E. {\it et al.}  [WMAP Collaboration],
  Astrophys.\ J.\ Suppl.\  {\bf 192} (2011) 18
  [arXiv:1001.4538 [astro-ph.CO]].
  
\bibitem[\protect\citeauthoryear{Kravtsov et al.}{2004}]{Kravtsov:2003sg} 
  Kravtsov~A.~V., Berlind~A.~A., Wechsler~R.~H., Klypin~A.~A., Gottloeber~S., Allgood~B. and Primack~J.~R.,
  Astrophys.\ J.\  {\bf 609}, 35 (2004)
  [astro-ph/0308519].
  
\bibitem[\protect\citeauthoryear{Kroupa}{2001}]{Kroupa:2000iv} 
  Kroupa~P.,
  Mon.\ Not.\ Roy.\ Astron.\ Soc.\  {\bf 322}, 231 (2001)
  [astro-ph/0009005].
  
\bibitem[\protect\citeauthoryear{Landy \& Szalay}{1993}]{Landy:1993yu}
  Landy~S.~D. and Szalay~A.~S.,
  Astrophys.\ J.\  {\bf 412} (1993) 64.
  
\bibitem[\protect\citeauthoryear{Leauthaud et al.}{2011}]{Leauthaud:2011zt} 
  Leauthaud~A., Tinker~J., Behroozi~P.~S., Busha~M.~T. and Wechsler~R.,
  Astrophys.\ J.\  {\bf 738}, 45 (2011)
  [arXiv:1103.2077 [astro-ph.CO]].
  
\bibitem[\protect\citeauthoryear{Leauthaud et al.}{2012}]{Leauthaud:2011rj} 
  Leauthaud~A., {\it et al.},
  Astrophys.\ J.\  {\bf 744}, 159 (2012)
  [arXiv:1104.0928 [astro-ph.CO]].
  
\bibitem[\protect\citeauthoryear{Lee et al.}{2006}]{Lee:2005jha}
  Lee~K., Giavalisco~K., Gnedin~O.~Y., Somerville~R., Ferguson~H., Dickinson~M. and Ouchi~M.,
  Astrophys.\ J.\  {\bf 642} (2006) 63
  [astro-ph/0508090].
  
\bibitem[\protect\citeauthoryear{Lewis \& Bridle}{2002}]{Lewis:2002ah}
  Lewis~A. and Bridle~S.,
  Phys.\ Rev.\ D {\bf 66} (2002) 103511
  [astro-ph/0205436].
  
\bibitem[\protect\citeauthoryear{Li et al.}{2006}]{Li:2005er}
  Li~C., Kauffmann~G., Jing~Y.~P. and White~S.~D.~M.,
  Mon.\ Not.\ Roy.\ Astron.\ Soc.\  {\bf 368} (2006) 21
  [astro-ph/0509873].
  
\bibitem[\protect\citeauthoryear{Liu et al.}{2010}]{Liu:2009rm}
  Liu~L., Yang~X., Mo~H.~J., van den Bosch~F.~C. and Springel~V.,
  Astrophys.\ J.\  {\bf 712} (2010) 734
  [arXiv:0912.1257 [astro-ph.GA]].
  
\bibitem[\protect\citeauthoryear{Longhetti \& Saracco}{2009}]{Longhetti:2008gv} 
  Longhetti~M. and Saracco~P.,
  Mon.\ Not.\ Roy.\ Astron.\ Soc.\  {\bf 394} 774 (2009)
  arXiv:0811.4041 [astro-ph].
  
\bibitem[\protect\citeauthoryear{Ma \& Fry}{2000}]{Ma:2000ik} 
  Ma~C.~-P. and Fry~J.~N.,
  Astrophys.\ J.\  {\bf 543}, 503 (2000)
  [astro-ph/0003343].
  
\bibitem[\protect\citeauthoryear{Magliocchetti \& Porciani}{2003}]{Magliocchetti:2003ee} 
  Magliocchetti~M. and Porciani~C.,
  Mon.\ Not.\ Roy.\ Astron.\ Soc.\  {\bf 346}, 186 (2003)
  [astro-ph/0304003].
  
\bibitem[\protect\citeauthoryear{Mandelbaum et al.}{2006}]{Mandelbaum:2005nx} 
  Mandelbaum~R., Seljak~U., Kauffmann~G., Hirata~C.~M. and Brinkmann~J.,
  Mon.\ Not.\ Roy.\ Astron.\ Soc.\  {\bf 368}, 715 (2006)
  [astro-ph/0511164].
  
\bibitem[\protect\citeauthoryear{Matsuoka et al.}{2011}]{Matsuoka:2010ba} 
  Matsuoka~Y., Masaki~S., Kawara~K. and Sugiyama~N.,
  Mon.\ Not.\ Roy.\ Astron.\ Soc.\  {\bf 410}, 548 (2011)
  [arXiv:1008.0516 [astro-ph.CO]].
  
\bibitem[\protect\citeauthoryear{Meneux et al.}{2007}]{Meneux:2007jr} 
  Meneux~B. {\it et al.},
  [arXiv:0706.4371 [astro-ph]].
  
\bibitem[\protect\citeauthoryear{More et al.}{2009}]{More:2008za} 
  More~S., van den Bosch~F.~C., Cacciato~M., Mo~H., Yang~X. and Li~R.,
  Mon.\ Not.\ Roy.\ Astron.\ Soc.\  {\bf 392}, 801 (2009)
  [arXiv:0807.4532 [astro-ph]].
  
\bibitem[\protect\citeauthoryear{Navarro, Frenk \& White}{1996}]{Navarro:1996gj}
  Navarro~J.~F., Frenk~C.~S. and White~S.~D.~M.,
  Astrophys.\ J.\  {\bf 490}, 493 (1997)
  [arXiv:astro-ph/9611107].
  
\bibitem[\protect\citeauthoryear{Peacock \& Smith}{2000}]{Peacock:2000qk} 
  Peacock~J.~A. and Smith~R.~E.,
  Mon.\ Not.\ Roy.\ Astron.\ Soc.\  {\bf 318}, 1144 (2000)
  [astro-ph/0005010].
  
 \bibitem[\protect\citeauthoryear{Peebles}{1973}]{Peebles:1973} 
 Peebles~P.~J.~E.,
ApJ, 185, 413 (1973)

\bibitem[\protect\citeauthoryear{Peebles}{1974}]{Peebles:1974} 
Peebles~P.~J.~E., 
AJ, 189, L51 (1974) 
  
\bibitem[\protect\citeauthoryear{Porciani, Magliocchetti \& Norberg}{2007}]{Porciani:2004vi} 
  Porciani~C., Magliocchetti~M. and Norberg~P.,
  Mon.\ Not.\ Roy.\ Astron.\ Soc.\  {\bf 355}, 1010 (2004)
  [astro-ph/0406036].
  
\bibitem[Scoccimarro et al.(2001)]{2001ApJ...546...20S} Scoccimarro, R., 
Sheth~R.~K., Hui~L., \& Jain~B., 
ApJ, 546, 20 (2001)

\bibitem[\protect\citeauthoryear{Scoccimarro et al.}{2001}]{Scoccimarro:2001} 
Scoccimarro~R., Sheth, R.~K., Hui, L., \& Jain, B.
ApJ, 546, 20 (2001)
  
\bibitem[\protect\citeauthoryear{Seljak}{2000}]{Seljak:2000gq} 
  Seljak~U.,
  Mon.\ Not.\ Roy.\ Astron.\ Soc.\  {\bf 318}, 203 (2000)
  [astro-ph/0001493].
  
\bibitem[\protect\citeauthoryear{Simha et al.}{2009}]{Simha:2008hd} 
  Simha~V., Weinberg~D.~H., Dave~R., Gnedin~O.~Y., Katz~N. and Keres~D.,
  Mon.\ Not.\ Roy.\ Astron.\ Soc.\  {\bf 399}, 650 (2009)
  [arXiv:0809.2999 [astro-ph]].
  
\bibitem[\protect\citeauthoryear{Smith et al.}{2003}]{Smith:2002dz}
  Smith~R.~E. {\it et al.}  [The Virgo Consortium Collaboration],
  MNRAS {\bf 341} (2003) 1311
  [arXiv:astro-ph/0207664].
  
\bibitem[\protect\citeauthoryear{Snaith et al.}{2011}]{Snaith:2011fy}
  Snaith~O.~N., Gibson~B.~K., Brook~C.~B., Courty~S., Sanchez-Blazquez~P., Kawata~D., Knebe~A. and Sales~L.~V.,
  arXiv:1104.2447 [astro-ph.GA].
  
\bibitem[\protect\citeauthoryear{Spergel et al}{2003}]{Spergel:2003cb}
  Spergel~D.~N. {\it et al.}  [WMAP Collaboration],
  Astrophys.\ J.\ Suppl.\  {\bf 148} (2003) 175
  [astro-ph/0302209].
  
\bibitem[\protect\citeauthoryear{Springel et al}{2001}]{Springel:2000qu} 
  Springel~V., White~S.~D.~M., Tormen~G. and Kauffmann~G.,
  Mon.\ Not.\ Roy.\ Astron.\ Soc.\  {\bf 328}, 726 (2001)
  [astro-ph/0012055].
  
\bibitem[\protect\citeauthoryear{Springel et al}{2005}]{Springel:2005nw} 
  Springel~V. {\it et al.},
  Nature {\bf 435}, 629 (2005)
  [astro-ph/0504097].
  
\bibitem[\protect\citeauthoryear{Tinker et al.}{2005}]{Tinker:2004gf}
  Tinker~J.~L., Weinberg~D.~H., Zheng~Z. and Zehavi~I.,
  Astrophys.\ J.\  {\bf 631}, 41 (2005)
  [arXiv:astro-ph/0411777].
  
\bibitem[\protect\citeauthoryear{Tinker et al.}{2005}]{Tinker:2005na}
  Tinker~J.~L., Weinberg~D.~H. and Zheng~Z.,
  Mon.\ Not.\ Roy.\ Astron.\ Soc.\  {\bf 368}, 85 (2006)
  [arXiv:astro-ph/0501029].
  
\bibitem[\protect\citeauthoryear{Tinker et al.}{2006}]{Tinker:2006sk} 
  Tinker~J.~L., Norberg~P., Weinberg~D.~H. and Warren~M.~S.,
  Astrophys.\ J.\  {\bf 659}, 877 (2007)
  [astro-ph/0603543].
  
\bibitem[\protect\citeauthoryear{Tinker et al.}{2008}]{Tinker:2008ff}
  Tinker~J.~L. {\it et al.},
  Astrophys.\ J.\  {\bf 688}, 709 (2008)
  [arXiv:0803.2706 [astro-ph]].
  
\bibitem[\protect\citeauthoryear{Tinker et al.}{2010}]{Tinker:2010my}
  Tinker~J.~L., Robertson~B.~E., Kravtsov~A.~V., Klypin~A., Warren~M.~S., Yepes~G. and Gottlober~S.,
  arXiv:1001.3162 [astro-ph.CO].
  
\bibitem[\protect\citeauthoryear{Totsuji \& Kihara}{1969}]{Totsuji:1969} 
Totsuji, H., \& Kihara, T., 
PASJ, 21, 221 (1969) 
  
\bibitem[\protect\citeauthoryear{van den Bosch, Yang \& Mo}{2003}]{vandenBosch:2002zn}
  van den Bosch~F.~C., Yang~X. and Mo~H.~J.,
  Mon.\ Not.\ Roy.\ Astron.\ Soc.\  {\bf 340} (2003) 771
  [arXiv:astro-ph/0210495].
  
\bibitem[\protect\citeauthoryear{Wake et al.}{2011}]{Wake:2010um}
  Wake~D.~A., {\it et al.},
  Astrophys.\ J.\  {\bf 728} (2011) 46
  [arXiv:1012.1317 [astro-ph.CO]].
  
\bibitem[\protect\citeauthoryear{Watson et al.}{2011}]{Watson:2011cz} 
  Watson~D.~F., Berlind~A.~A. and Zentner~A.~R.,
  Astrophys.\ J.\  {\bf 738}, 22 (2011)
  [arXiv:1101.5155 [astro-ph.CO]].
  
\bibitem[\protect\citeauthoryear{Weinmann et al.}{2006}]{Weinmann:2006cq}
  Weinmann~S.~M., van den Bosch~F.~C., Yang~X., Mo~H.~J., Croton~D.~J. and Moore~B.,
  Mon.\ Not.\ Roy.\ Astron.\ Soc.\  {\bf 372} (2006) 1161
  [astro-ph/0606458].
  
\bibitem[\protect\citeauthoryear{White \& Frenk}{1991}]{White:1991mr} 
  White~S.~D.~M. and Frenk~C.~S.,
  Astrophys.\ J.\  {\bf 379}, 52 (1991).
  
\bibitem[\protect\citeauthoryear{Worthey}{1994}]{Worthey:1994iw}
  Worthey~G.,
  Astrophys.\ J.\ Suppl.\  {\bf 95} (1994) 107.
  
\bibitem[\protect\citeauthoryear{Yang et al.}{2005}]{Yang:2004qi}
  Yang~X.~-H., Mo~H.~J., Jing~Y.~P. and van den Bosch~F.~C.,
  Mon.\ Not.\ Roy.\ Astron.\ Soc.\  {\bf 358} (2005) 217
  [astro-ph/0410114].
  
\bibitem[\protect\citeauthoryear{Yang et al.}{2008}]{Yang:2007pg}
  Yang~X., Mo~H.~J. and van den Bosch~F.~C.,
  Astrophys.\ J.\  {\bf 676} (2008) 248
  [arXiv:0710.5096 [astro-ph]].
  
\bibitem[\protect\citeauthoryear{Zehavi et al.}{2005a}]{Zehavi:2004zn}
  Zehavi~I. {\it et al.} 2005a  [SDSS Collaboration],
  Astrophys.\ J.\  {\bf 621} (2005) 22
  [arXiv:astro-ph/0411557].
  
\bibitem[\protect\citeauthoryear{Zehavi et al.}{2005b}]{Zehavi:2004ii} 
  Zehavi~I. {\it et al.} 2005b  [The SDSS Collaboration],
  Astrophys.\ J.\  {\bf 630}, 1 (2005)
  [astro-ph/0408569].
  
\bibitem[\protect\citeauthoryear{Zehavi et al.}{2010}]{Zehavi:2010bh} 
  Zehavi~I. {\it et al.}  [The SDSS Collaboration],
  Astrophys.\ J.\  {\bf 736}, 59 (2011)
  [arXiv:1005.2413 [astro-ph.CO]].
  
\bibitem[\protect\citeauthoryear{Zheng et al.}{2005}]{Zheng:2004id}
  Zheng~Z. {\it et al.},
  Astrophys.\ J.\  {\bf 633} (2005) 791
  [arXiv:astro-ph/0408564].
  
\bibitem[\protect\citeauthoryear{Zheng, Coil \& Zehavi}{2007}]{Zheng:2007zg}
  Zheng~Z., Coil~A.~L. and Zehavi~I.,
  Astrophys.\ J.\  {\bf 667} (2007) 760
  [astro-ph/0703457 [ASTRO-PH]].
  
\bibitem[\protect\citeauthoryear{Zheng et al.}{2009}]{Zheng:2008np} 
  Zheng~Z., Zehavi~I., Eisenstein~D.~J., Weinberg~D.~H. and Jing~Y.,
  Astrophys.\ J.\  {\bf 707}, 554 (2009)
  [arXiv:0809.1868 [astro-ph]].

\end{thebibliography}
\end{document}